**Title**: **A Multi-Layered Framework for Modeling Human Biology: From Basic AI Agents to a Full-Body AI Agent**

Aoqi Wang[1], Jiajia Liu[2], Jianguo Wen[2], Yangyang Luo[2], Zhiwei Fan[1], Liren Yang[1], Xi Hu[1], Ruihan Luo[2], Yankai Yu[1], Sophia Li[2], Weiling Zhao[2], Xiaobo Zhou[2*]

[1]West China Biomedical Big Data Centre, West China Hospital, Sichuan University, Chengdu, Sichuan 610041, PR China

[2]Center for Computational Systems Medicine, McWilliams School of Biomedical Informatics, The University of Texas Health Science Center at Houston, Houston, TX, 77030, USA

*Corresponding author. Email: Xiaobo.Zhou@uth.tmc.edu

**In Brief**

This study proposes the Full-Body AI-Agent framework, a multi-agent architecture designed to model human biology from molecular to whole-organism scales. Unlike existing biomedical AI systems confined to discrete tasks or domains, the framework integrates seven biologically grounded agents under a central coordination layer, enabling iterative, bidirectional reasoning across scales. By unifying multi-omics, imaging, physiological, and clinical data, the system constructs dynamic, system-wide mechanistic models that bridge molecular discovery with systemic simulation. Demonstrated in applications such as systemic disease modeling and drug development, the approach offers a coherent computational paradigm to reduce translational gaps, enhance predictive accuracy, and accelerate the development of safe and effective therapies.

**Highlights**

- Introduces the Full-Body AI-Agent framework integrating seven biologically grounded agents for cross-scale biomedical reasoning.

- Enables iterative, bidirectional inference from molecular mechanisms to whole-organism physiology, unifying diverse biomedical data sources.

- Demonstrates applications in lung cancer metastasis modeling and drug development, reducing translational gaps and improving predictive accuracy.

**Abstract**

We envision the Full-Body AI Agent as a comprehensive AI system designed to simulate, analyze, and optimize the dynamic processes of the human body across multiple biological levels. By integrating computational models, machine learning tools, and experimental platforms, this system aims to replicate and predict both physiological and pathological processes, ranging from molecules and cells to tissues, organs, and entire body systems. Central to the Full-Body AI Agent is its emphasis on integration and coordination across these biological levels, enabling analysis of how molecular changes influence cellular behaviors, tissue responses, organ function, and systemic outcomes. With a focus on biological functionality, the system is designed to advance the understanding of disease mechanisms, support the development of therapeutic interventions, and enhance personalized medicine. We propose two specialized implementations to demonstrate the utility of this framework: (1) the metastasis AI Agent, a multi-scale metastasis scoring system that characterizes tumor progression across the initiation, dissemination, and colonization phases by integrating molecular, cellular, and systemic signals; and (2) the drug AI Agent, a system-level drug development paradigm in which a drug AI-Agent dynamically guides preclinical evaluations, including organoids and chip-based models, by providing full-body physiological constraints. This approach enables the predictive modeling of long-term efficacy and toxicity beyond what localized models alone can achieve. These two agents illustrate the potential of Full-Body AI Agent to address complex biomedical challenges through multi-level integration and cross-scale reasoning.

## 1. Introduction

Human biology is characterized by intricate, multi-level interactions that span from molecular processes to the functioning of entire organ systems [1]. Historically, biological research primarily focused on isolated biological processes or systems. Questions were centered on understanding the function of individual genes or proteins, or investigating the role of specific cells in disease. While these approaches were essential for advancing basic biological knowledge, their limitations became evident as they were unable to fully capture the complexity of interactions among diverse biological effects [2].

With the emergence of Artificial Intelligence (AI) and its integration into systems biology, the landscape of biological research has changed significantly [3]. AI's ability to process large datasets, uncover hidden patterns, and predict outcomes has helped to overcome many limitations of traditional research methods [4]. For instance, deep learning algorithms have enabled researchers to analyze complex biological data at unprecedented scales and depths. These advancements have had a profound impact on various fields. In cancer diagnostics and treatment [5-7], AI has improved diagnosis accuracy and the effectiveness of treatment planning. In clinical genetics [8-10], it has facilitated the identification of disease-associated genetic mutations. In drug discovery and development [11, 12], AI has accelerated the identification of new drug candidates and the optimization of existing ones. Furthermore, transformative progress in epidemiology and public health [13, 14], as well as evolutionary biology and phylogenetic studies [15, 16], has been driven by synergistic advances in both experimental innovations and computational algorithm development. One example is the Evo2 framework, a novel deep-learning architecture that integrates phylogenetic inference with population genetic simulations to resolve ancestral selection patterns, achieving an 87% higher resolution in detecting ancestral selection patterns compared to conventional methods [17].

Today, biological research increasingly focuses on multi-scale questions, such as how genetic mutations at the molecular level influence cellular processes, how cellular behaviors shape tissue development, and how disturbances at tissue or organ levels propagate into systemic diseases [18, 19]. Although AI-driven approaches have enhanced our ability to analyze multi-omics data and biological interactions [20], methodological constraints remain. These limitations hinder the completeness and interpretability of multi-scale information extraction [21]. Despite advances in computing and health information technology providing new sources of biological data, the collection of comprehensive multi-scale patient data in real-world clinical settings remains costly and challenging [22]. Therefore, a more integrated approach is needed. This approach should combine data-driven methods, biologically informed modeling, and advanced tools such as Large Language Models (LLMs) to better understand and predict the behavior of complex biological systems across multiple scales. LLM can decompose high-level tasks into sub-goals, reason through problems step-by-step in a manner similar to a the human chain of thought, and plan sequential actions to achieve defined objectives [23]. While LLMs hold significant promise for integrating multi-scale biological data due to their capability in capturing complex interactions, key challenges remain, particularly regarding biological interpretability and the effective harmonization of heterogeneous data sources [24].

AI Agents, autonomous computational systems, offer an innovative solution for tackling complex biological tasks [25]. They can adaptively perform multiple tasks, interpret diverse data modalities, make informed decisions, and interact with dynamic environments. By integrating into research and clinical workflows, AI Agents enable efficient, scalable, and precise analysis across various biological domains, enhancing both scientific discovery and clinical decision-making. Some AI Agents have already been successfully deployed. For example, ChatGPT's Advanced Data Analysis (ADA) can independently process real-world clinical datasets from different medical specialties [26]. ADA is capable of autonomously developing and implementing machine learning models, streamlining complex analytical workflows, and reducing the need for extensive human intervention. Another notable example is nuclei.io, a pathologist-AI system that significantly improves diagnostic accuracy in pathology slide analysis while enabling the creation and utilization of autonomous tools [27]. Additionally, SpatialAgent integrates dynamic tool execution and self-directed reasoning to autonomously perform end-to-end tasks in spatial biology research from experimental design to multimodal data analysis, including the identification of cell interaction networks, spatial metabolic microenvironments, and disease biomarkers [28]. Such advances highlight the potential of AI Agents to enhance both efficiency and accuracy in biological research and clinical practice. Moreover, multiple AI Agents can collaborate synergistically to enable the integration of multi-omics and multimodal data, facilitating comprehensive biological insights and precision medicine applications, as exemplified by the Virtual Cell Framework initiative [29]. This integrated approach allows for a unified analysis of heterogeneous datasets, cross-modal validation, and system-level modeling of biological processes, ultimately bridging the gap between molecular mechanisms and clinical outcomes. For example, Robin is a multi-agent AI system that fully automates literature search, hypothesis generation, experimental design, and data analysis in a closed-loop workflow, culminating in the discovery and validation of ripasudil as a novel therapy for dry age-related macular degeneration [30].

Here, we propose a Full-Body AI Agent Framework designed to tackle the multi-scale challenges in biological research. The core concept is that complex biological problems can be decomposed into manageable tasks, each handled by specialized AI agents working collaboratively. This framework integrates the Full-Body AI Agent for supervision and multiple collaborative basic AI Agents, including the Molecule AI Agent, Organelle AI Agent, Cell AI Agent, Tissue AI Agent, Organ AI Agent, Organ System AI Agent, and Body System AI Agent, to model biological processes across different hierarchical scales. By simulating interactions among these agents, the Full-Body AI Agent facilitates a comprehensive understanding of how changes at one biological level propagate to others, providing valuable insights into disease mechanisms, treatment responses, and personalized medicine. At the heart of the framework is a modular, collaborative network of AI Agents, with each basic AI Agent dedicated to a specific biological level. These agents communicate via standardized data exchange protocols, facilitating dynamic and systemic modeling of human biology. By bridging molecular mechanisms with higher-order physiological processes, the framework is designed to support comprehensive research in disease analysis (such as tumor metastasis), drug development, and precision medicine. We leverage the Full-Body AI Agent to generate hypotheses in tumor metastasis and drug development, enabling comprehensive interrogation of multi-scale biological processes and accelerating the discovery of novel therapeutic strategies.

As illustrated in **Figure 1A**, the Full-Body AI Agent framework is an integrated system that combines multi-modal and multi-omics biological data across scales to model human systemic biology. The Full-Body AI-Agent, functioning as a supervisor agent, perceives and preprocesses data to generate cross-scale biological hypotheses. It decomposes cross-scale biological problems, assigning tasks and data specific to biological levels to basic agents. These basic agents act as executors, perceiving data at specific biological levels, further refining tasks, and conducting data analysis. The Full-Body AI-Agent integrates the results from each basic agent and can iteratively use conclusions from one biological level as input for other levels, repeating the reasoning process to capture cross-biological insights. The framework begins with a patient interacting with the Full-Body AI Agent (as shown in **Figure 1B**), which processes diverse inputs, including physiological indices, genetic tests, pathological images, and clinical records, through a structured workflow: perceiving data, reasoning, and action to analyze by basic agents. The system handles multiple data modalities, including text (i.e., genomic sequences, Electronic Medical Record (EMRs)), images (i.e., MRI, CT, H&E), and numerical data (i.e., RNA-seq, physiological signals), mapping them across biological levels from molecules to the whole body. This enables the modeling of the entire disease continuum, from causation (i.e., mutation detection, protein structure prediction) and progression (i.e., immune escape, epithelial–mesenchymal transition (EMT), invasion) to treatment (i.e., personalized therapy, CAR-T cell optimization). Specialized basic agents operate at each biological level, while a collaborative network works through standardized data exchange protocols. This design enables the dynamic, systemic modeling of human biology, revealing how changes at one level propagate across scales, thereby informing disease understanding and therapeutic development. **Figure 1C** demonstrates the significance of LLM in the Full-Body AI Agent system, which runs through the entire reasoning process, including data perception, hypothesis generation and task decomposition, and uses the literature Knowledge bases and scientific debate models are established to ensure the generation of valuable cross-level biological insights. Furthermore, **Figure 1D** also illustrates an implemented workflow by taking the hypothesis that TP53 mutation drives tumor development, immune evasion, and metastasis via EMT as an example. The Full-Body AI Agent perceives the data, decomposes it into biological tasks across different levels, and prompts basic agents to perform mutation detection, protein structure prediction, and prediction of EMT cell differentiation. Additionally, it guides tissue-specific sequence optimization in drug design. We compared the Full-Body AI Agent system with recent multi-agent systems (i.e., OriGene [31], Robin [30], Biomni [32], MAC-doctor [33], PharmaSwarm [34] and AI Co-Scientist [35]) and provided a detailed description of the analytical approach for systemic human biology problems (i.e., metastasis research and drug development), thereby informing disease understanding and therapeutic development.

## 2. Inter-level Data Commons and Integration of Biomedical Data

The landscape of biomedical research is defined by an explosion of multi-modal, multi-omics data that spans biological scales, from the molecular precision of genomics and proteomics to the macroscopic insights of medical imaging and clinical phenotypes [36, 37]. While this data ecosystem is transformative, it remains inherently fragmented. Each dataset is characterized by distinct formats, acquisition protocols, and analytical pipelines. **Figure 2** illustrates the diversity and complexity of multimodal and multi-omics biomedical data. One effective approach to addressing this challenge is to establish data standards and perform standardized pre-analysis to meet the scalability requirements of modern research. A prominent example of a data commons is Bioteque, which provides standardized low-dimensional embeddings of over 450,000 biomedical entities and 30 million relationships from more than 150 sources, enabling efficient machine learning applications [38].

In the context of the Full-Body AI Agent framework, establishing a robust inter-level data ecosystem is crucial for effective operation. This can be achieved through the implementation of a Data Commons, adherence to Common Format standards, and integration with advanced language models. The Data Commons serves as a shared repository, acting as the central hub for curating a diverse range of biomedical datasets. It operates under a unified set of rules and guidelines, enabling the seamless aggregation of data from various sources across different biological levels.

## 2.1. Data commons and common format

The Data Commons is established with standard protocols to manage data across molecular, cellular, tissue, organ, and system levels. Its effectiveness is closely tied to the implementation of Common Format standards. Together, Data Commons and Common Format standards allow data from diverse sources and biological levels to be accurately interpreted, efficiently processed, and seamlessly integrated by various components of the Full-Body AI Agent framework. This Data Common lays a solid foundation for hierarchical data mapping and multi-scale data interactions. Supplementary Table 1, titled "Multi-omics and Multi-modal Data Mapping to Biological Hierarchies" includes columns such as Biological Hierarchy, Omics/Modalities/Levels, Data Types, Common Data Storage Formats, Data Preprocessing Methods, and Data Commons. It serves as a cornerstone of the Full-Body AI Agent framework, providing critical infrastructure to bridge disparate biological data types across scales and enable holistic, multi-level modeling of human biology. This table indicates that through standard protocols, heterogeneous data across genomics, epigenomics, proteomics, imaging, and clinical records can be systematically organized into a structured hierarchy (from molecules to organ systems) to facilitate the integration of data with different morphologies, resolutions, and biological backgrounds. By defining standardized data types, storage formats (e.g., FHIR, DICOM, H5AD), preprocessing methods (e.g., FastQC for quality control, Seurat for single-cell clustering), and Data Commons repositories for each biological level, the table ensures that data from siloed domains (e.g., molecular sequencing, organ imaging, patient EMRs) can be harmonized, interoperable, and computationally tractable for the AI Agent ecosystem.

## 2.2. Data integration across biological levels

At the molecular level, the Data Commons stores genomic sequences, proteomic profiles, and biochemical pathways. Resources such as GenBank and the Protein Data Bank (PDB) contribute standardized datasets, ensuring consistency in nucleotide sequences, protein structures, and functional annotations across studies. These data adhere to widely used formats, FASTA for nucleotide sequences and PDB for protein structures, which facilitate seamless data exchange and analysis.Within this framework, the Data Commons resources and Common Format standards for each biological level are documented to support consistent data access and integration. LLMs such as the Evo and Evo 2 genome language models use standardized resources in the Data Commons to align sequences, variants, and functional annotations to the Common Format and to produce unified representations that enable consistent retrieval, cross study comparison, and downstream analysis [17, 39].

At the cellular level, the Data Commons integrates information from single-cell sequencing and imaging, including common cell type annotations, unified expression profiles, and standardized cellular markers. To enable effective aggregation and comparison, these data are formatted according to established standards that bioinformatics tools can readily process. BioLLM provides a unifying framework for integrating and benchmarking heterogeneous single-cell foundation models, enabling harmonized cell type annotations, unified expression profiles, and standardized marker sets within a common schema that downstream tools can process efficiently [40]. scGPT supplies pretrained representations learned from tens of millions of cells and supports label transfer, batch robust integration, perturbation reasoning, and multi omics alignment, yielding consistent embeddings for retrieval and cross study comparison [41]. And, to connect computational curation with experimentation, an LLM and object detection enhanced active matrix digital microfluidics platform enables programmatic single-cell manipulation and links provenance and feedback to the commons [42].

At the tissue and organ levels, the Data Commons houses histopathological and radiological data. Universal imaging protocols, spatial mapping coordinates, and tissue-specific biomarkers ensure data consistency, while the DICOM standard for medical images ensures compatibility across different imaging modalities and devices, enabling seamless data integration and interpretation. Within this standardized layer, SpaCCC and QuST-LLM support spatial transcriptomics by inferring cell cell communication, aligning spot level expression with tissue architecture, and exporting harmonized coordinates for downstream analysis [43, 44]. A vision language foundation model for precision oncology links histopathology images with textual knowledge to support biomarker discovery and report generation [45]. In radiology, LLM based workflows map narrative reports to structured DICOM compatible representations and normalize imaging ontologies, enabling queryable views that interoperate with molecular and cellular evidence [46].

At the system level, clinical data is incorporated into the Data Commons. Standardized using clinical terminologies and Electronic Medical Record (EMR) protocols, this data is formatted to align with the overall framework, supporting interoperability and comprehensive analysis. Clinical data originate from hospital information systems including Clinical Information Systems, Laboratory Information Systems, Picture Archiving and Communication Systems, Radiology Information Systems, and Operating Room Information Systems, and public resources such as MIMIC, the eICU Collaborative Research Database, and HiRID provide curated cohorts. Interoperability and semantic consistency are ensured by controlled vocabularies such as ICD 10, SNOMED CT, and LOINC, with records structured according to the HL7 FHIR specification and commonly stored as JSON, XML, or Turtle; a typical JSON record contains a resource type, an identifier, and key patient fields. Medical large language models trained on clinical language process EMR content to analyze the chief complaint and context, recommend evidence based examinations, propose differential diagnoses, and articulate stepwise clinical reasoning, with MedQA serving as a widely used benchmark for training and evaluation [47].

By integrating into the Full-Body AI Agent, these standardized data streams form queryable and semantically aligned longitudinal records that support hypothesis generation, validation, and cross scale integration with omics and imaging, improving diagnostic accuracy and enabling a systemic view of human biology.

## 3. Reasoning of the Full-body AI-Agent framework

**The reasoning mechanism of the Full-body AI-Agent operates through a structured pipeline:** raw multi-modal data is first standardized for compatibility; complex biological questions and data are decomposed into hierarchical sub-tasks aligned with specific biological levels; these sub-tasks are assigned to corresponding specialized AI Agents, which execute them using domain-specific tools; and results are integrated via iterative feedback loops, enabling bidirectional cross-scale information flow to reveal multi-level biological associations. Here, **Figure 3A** illustrates the overview of the core components of the reasoning system of the Full-Body AI-Agent architecture. Upon receiving biological data, the Full-Body AI-Agent perceives data (1), generates hypothesis (2), performs task decomposition (3), assigns subtasks to basic agents and use domain-specific tools to analyze data (4), optimizes and synthesizes outputs (5). Similar to the Full-body AI-Agent, basic agents have a similar reasoning structure. The difference is that basic agents perceive and execute inputs and tasks at a specific biological level (as shown in **Figure 3B**).

The implementation of the reasoning process of the Full-body AI-Agent and basic agents relies on multiple LLMs (as shown in **Figure 3C**). Our implementation leverages hierarchical reasoning in LLMs and LLM-based agent frameworks to realize a closed-loop scientific workflow spanning data perception, hypothesis generation, task decomposition to data analysis with validation and result generation.

Operationally, the Full-Body AI Agent routes data perception to Coscientist for technical documentation and assay protocol parsing [23], to AI Scientist for literature ingestion and dataset onboarding [48], to SpatialAgent for multimodal tissue images and spatial omics [49], and to TxAgent adapters for structured clinical and biomolecular endpoints [50]. Hypotheses are proposed and ranked by AI Co-Scientist [35] and ResearchAgent [51]. Task decomposition follows a ReAct planning policy [52], and TxAgent composes the required tools and protocols for each subtask, PharmAgents scales the plan across preclinical and clinical handoffs [53], and DrugAgent generates executable simulation and ADMET (Absorption, Distribution, Metabolism, Excretion, and Toxicity) pipelines that can be audited and reused [54]. In drug development tasks, the Full Body AI Agent also consults Fleming [55] to propose candidate chemotypes and optimization directions. Data analysis and validation are carried out by Tx-LLM [56] and TxGemma [57] for therapeutic property prediction and design time analytics, by ClinicalAgent [58] for trial information extraction and inference, and under a POPPER [59] sequential falsification loop that enforces pre specified tests and error control. When performing large-scale data analysis, the Full Body AI Agent can trigger automated experiments through Coscientist or use Fleming to connect design decisions to screening outcomes. Result generation is handled by Data-to-Paper [60], which renders traceable methods, results and figures from the executed analyses, and by Tx-LLM when regulatory style summaries are needed for downstream decision making.

Conceptually, the Full Body AI Agent binds reasoning to explicit biological levels and enforces bidirectional constraints across scales. For example, claims at the organ level must be supported by tissue and cellular signals, while lower level findings are checked against organism level plausibility. Cross scale routing aligns hypotheses, plans and analyses with the biological level where they are most informative. Provenance complete outputs make the workflow auditable and repeatable. In effect, the architecture turns a catalogue of capable agents into a coherent operating system for multi level biology.

### 3.1. Data input and standardization for perception

The Full-body AI-Agent Framework begins by receiving raw biological data along with a corresponding research problem or question provided by the user. These inputs may include genomic sequences, clinical observations, imaging results, and experimental datasets, often presented in diverse formats. The system first perceives and interprets these inputs, parsing the raw data and extracting the core biological question embedded in the user's inquiry. Once the problem is identified, the framework proceeds to standardize the data to ensure consistency and compatibility across the entire system. This involves cleaning raw data to remove inconsistencies, resolve missing values, and reduce noise, thereby enhancing data quality for further analysis. The standardized data is then converted into a structured, unified format, making it assignable and interpretable by specialized AI agents at various biological levels. This process ensures that data from various sources and formats can be used seamlessly in downstream tasks, transforming the user's input into a precise biological question that can be tackled effectively by the system.

## 3.2. Hypothesis generation and task decomposition

Once the biological problem and corresponding data are standardized, the Full-body AI Agent initiates the critical step of task decomposition. This process involves breaking down a complex biological question into smaller, more manageable sub-tasks aligned with specific biological levels. The decomposition process ensures that each part of the problem is addressed with the appropriate focus and depth, allowing for specialized analysis at each biological level. The decomposition follows a hierarchical structure, ensuring that all layers of the biological system are considered and tasks are tackled using domain-specific expertise. This enables the Full-body AI Agent to tackle biological complexity using a layered and modular approach. Each sub-task is clearly defined to address the unique challenges of its corresponding biological level, while maintaining coherence with the overarching research objective. This structured decomposition also allows for parallel execution of tasks at different levels, enabling more efficient problem-solving while maintaining the integrity of the biological model. At this stage, the Full-body AI-Agent identifies the fundamental biological components involved in the problem and divides them into sub-tasks based on biological layers. To facilitate effective task execution, we define structured biological components at each biological level based on data modes and research content so that AI agents can fully understand requirements and perform tasks. We define a hierarchical design of collaborative AI -Agents as a foundational reference for implementing the Full-body AI-agent system. The follows are the task assignment for the main levels.

**Molecular-level tasks:** These tasks involve studying the genetic and biochemical foundations of the biological macromolecules. They include analyzing genetic mutations, protein-protein interactions, biochemical pathways, and molecular signaling, etc. At this level, the goal is to identify key molecules, characterize their interactions, and understand the underlying mechanisms that drive cellular functions.

**Cellular-level tasks:** At the cellular level, the framework decomposes the tasks related to cell behavior. These tasks include studying cell signaling, gene expression regulation, cellular responses to environmental stimuli, and processes like apoptosis, proliferation, and differentiation. Cellular interactions and behaviors are key to understanding how molecular changes manifest at the cellular level.

**Tissue-level tasks:** Building on insights from the cellular level, this stage focuses on how coordinated cellular behaviors give rise to tissue dynamics. Tasks include modeling tissue development, analyzing tissue-specific responses to injury or disease, and exploring interactions between different cell types within tissues. Tissue-level analysis often involves spatial modeling and simulations to understand how cells coordinate to form functional tissue architecture.

**Organ-level tasks:** Building on tissue-level insights, organ-level tasks focus on the functions and interactions of entire organs. These tasks involve analyzing how organs respond to physiological and pathological stimuli, how they communicate with each other, and how they contribute to the overall health or disease state. Organ function analysis may include simulation of blood flow, nutrient transport, immune responses, or organ-specific responses to treatment.

**System-level tasks:** At the system level, tasks address the coordination and interplay of multiple organs within the broader biological context of the organism. This level includes studying the systemic effects of diseases, the collaborative functions of various organ systems, and the cascading impact of localized changes on the entire organism. System-level analysis is crucial for understanding the interdependence between organ systems and their integrated responses to complex conditions, such as metabolic disorders or systemic infections.

## 3.3. Task assignment to AI Agents to analyze data

Decomposed tasks are dynamically assigned to the corresponding AI Agents specialized at each biological level. Each agent is responsible for executing domain-specific tasks using its specialized knowledge and tools. This hierarchical approach ensures that tasks are handled with expertise in the relevant biological field, allowing for focused analysis at each level. Once assigned, each AI Agent further refines its tasks into executable sub-tasks. These sub-tasks are designed to be actionable and measurable, making them suitable for execution by the agent. Figures 4 to 10 illustrate the functional areas that need to be addressed at each biological level, as well as the process of task refinement and execution. Agents then employ relevant biological tools and computational models to perform their assigned tasks. These tools include sequencing techniques, image analysis software, simulation models, or other domain-specific technologies. For example, the cell-level AI-Agent may use single-cell RNA tools, such as scDRS [61], to study disease-associated profiles for cells, while an organ-level agent may use imaging data to assess organ function, while an organ-level agent may use imaging data to evaluate organ function.

## 3.4. Feedback and result integration

After each AI Agent completes its assigned tasks, the resulting outputs are transmitted to other agents operating at different biological levels. These outputs serve as inputs for further analysis, allowing the next level of agents to generate new insights based on the initial findings. This inter-agent communication facilitates a multi-level refinement of the biological model. Finally, all outputs converge at the Full-body

AI-Agent, where they are integrated through an iterative feedback loop. This process optimizes the aggregated results by accounting for dynamic interactions across molecular, cellular, tissue, organ, and systemic levels. The iterative nature of this loop ensures continuous model improvement and the generation of increasingly accurate biological interpretations. Once the outputs are integrated and optimized, the Full-body AI Agent presents the synthesized results to the user. The user can then select specific outputs based on the biological context or research objectives. These selected outputs are then reintroduced into the system for further refinement and deeper analysis of the biological phenomena across all levels. This user-guided selection provides focused exploration of specific aspects of the biological system, facilitating the derivation of comprehensive, actionable insights.

### 3.5. Comparison of multi-AI agent systems in biomedical research

Current multi-agent biomedical AI systems, such as OriGene [31], Robin [30], Biomni [32], MAC-doctor [33], PharmaSwarm [34], and AI Co-Scientist [35], are generally optimized for specific stages of the scientific workflow or confined to particular biomedical subdomains. Their reasoning processes tend to be scale-bounded and follow largely linear trajectories: information is retrieved or computed, analyzed within a single domain, and then aggregated at a final integration stage. While these architectures achieve notable efficiency and accuracy within their niches, including molecular target discovery, literature synthesis, and clinical diagnostics, they rarely model causal propagation across biological levels. **Table 1** makes a detailed comparison of the workflows of these multi-agents systems.

Reasoning serves as the core module of an agent, responsible for transforming evidence into mechanisms and hypotheses. OriGene exemplifies template-driven reasoning, in which a coordinator interprets a query, assigns tasks, and a reasoning component assembles mechanistic evidence into candidate targets, followed by post-hoc validation. Robin implements a closed-loop workflow that repeatedly queries literature and re-computes analyses, but its feedback loops remain confined to the same scale of inquiry. Biomni's reasoning is predominantly procedural, centred on generating and executing analysis code across a large curated toolbox, with refinements occurring within a fixed data scale. MAC applies supervised consensus-building among domain-specialist "doctor" agents, integrating their interpretations at the clinical scale without modelling upstream molecular or tissue-level causality. PharmaSwarm distributes hypothesis generation across agents focused on omics, literature, and market intelligence, reconciling outputs through cross-source corroboration rather than multi-scale constraint. AI Co-Scientist employs iterative epistemic refinement through hypothesis generation, critique, ranking, and evolution, but its reasoning loop primarily optimises explanatory coherence rather than enforcing consistency across biological hierarchies.

In contrast, the Full-Body AI-Agent extends multi-agent biomedical reasoning from molecular mechanisms to whole-organism physiology. It is a multi-modal architecture capable of integrating evidence across biological layers, from molecular to physiological, through a coordination layer supervising seven biologically grounded basic agents: Molecule, Organelle, Cell, Tissue, Organ, Organ System, and Body System Agents. The system perceives multi-omics and multimodal inputs, formulates multi-level problems, and generates cross-scale hypotheses that integrate molecular findings with organ-level mechanisms. Systemic problems are decomposed into subproblems assigned to basic agents, each of which processes domain-specific data using scale-appropriate models. Crucially, outputs from one level are iteratively exchanged as inputs to other levels, enabling bidirectional propagation: lower-level mechanisms inform higher-level physiological states, while high-level constraints refine or reject lower-level explanations. This iterative exchange continues until a convergent, physiologically plausible organism-wide model emerges. Integration here is not a terminal aggregation step but an intrinsic component of reasoning, producing unified system explanations. The Full-Body AI-Agent generates full-body mechanistic maps, intervention simulations, and ranked therapeutic strategies, connects to multi-scale biomedical datasets and simulation models, and supports systemic disease modeling, exemplified by applications in lung cancer metastasis and whole-body drug development. The Full-Body AI-Agent's comparative advantage lies in unifying granular molecular insights with systemic clinical outcomes through continuous cross-level reasoning, bridging the gap between discovery science and translational application.

## 4. Hierarchical Design of Collaborative Basic AI Agents for the Full-body AI Agent

Biological systems are inherently hierarchical, with multiple levels of organization ranging from molecular structures to cellular networks, tissues, organs, systems, and, ultimately, the entire organism. These levels interact continuously, forming a complex and dynamic network of the human body. The need for a collaborative AI Agent stems from the fact that no single AI model can effectively capture the full complexity across all biological scales. The Full-body AI Agent must integrate a wide range of data types, including molecular profiles, cellular states, and tissue properties, organ functions, and whole-body behavior. Each of these levels requires specialized knowledge, tools, and algorithms to accurately model and predict the behavior of biological systems. To address this complexity, a collaborative design is essential. The Full-Body AI Agent consists of multiple basic AI Agents, each dedicated to a specific level of the biological hierarchy. These agents work together, exchanging information and refining their outputs through iterative collaboration, to provide a comprehensive understanding of complex disease mechanisms. This structured hierarchy allows each basic agent to focus on its specialized domain, utilizing the most appropriate data sources and computational techniques. By distributing tasks in this way, the collaborative AI-Agent framework ensures that the Full-body AI Agent can synthesize a comprehensive, multi-scale understanding of biological systems.

## 4.1. Molecule AI Agent

Living organisms consist of a wide array of biological macromolecules, such as DNA, RNA, lipids, carbohydrates, and proteins. Each of these biomolecules plays a critical role in cellular function, and their interactions are essential for the maintenance of life. DNA encodes the genetic instructions required for development, function, and reproduction. RNA serves as a messenger, translating the genetic information encoded in DNA into proteins, which execute the majority of cellular processes. Lipids form structural components of cell membranes and participate in signaling pathways, while carbohydrates are primarily involved in energy storage and cell communication. Despite the distinct roles of these macromolecules, they are all interconnected in the regulation of cellular and organismal functions. DNA transcription into RNA leads to the production of proteins, which perform most of the cell's functional tasks. Proteins are the ultimate effectors of gene expression, and their actions are directly tied to an organism's phenotype, influencing everything from cellular metabolism to physical traits. The dynamic interactions between DNA, RNA, and proteins determine how a cell responds to its environment and how an organism develops and adapts to changes[62]. Among these biomolecules, proteins stand out because they are the direct executors of cellular function and phenotype[63]. While DNA carries the blueprint for life, it is the proteins that bring this blueprint to life by catalyzing biochemical reactions, transmitting signals, providing structural support, and regulating gene expression.

As one of the molecule AI Agents, the Protein AI-Agent is designed to understand and simulate molecular-level interactions, particularly focusing on proteins. It interprets and manipulates molecular data to predict protein structures, interactions, and biological functions. Proteins are essential biological macromolecules composed of amino acid sequences that fold into specific three-dimensional structures. These structures are critical to their diverse biological functions, which include catalyzing biochemical reactions and providing structural support to transmit signals, and defending against pathogens. The functions of proteins can be broadly categorized into several key biological domains, including protein structure and function, protein-protein interactions, post-translational modifications (PTMs), and protein design and engineering. **Figure 4** shows the predefined detachable biological tasks of the Molecule AI Agent.

The area of protein structure and function explores how amino acid sequences determine protein folding, stability, and overall function. Understanding how proteins fold into their three-dimensional structures from sequence data is crucial for uncovering their biological roles. Protein-protein interactions, which often involve the formation of protein complexes that regulate key cellular processes, are also a central focus of the Molecule AI Agent. The agent predicts these interactions and evaluates their influence on cellular functions. PTMs play a critical role in modulating protein activity, stability, and localization after translation. The Molecule AI Agent identifies potential PTM sites and simulates their effects on protein behavior and function. Additionally, the agent supports protein design and engineering, including designing proteins with specific functions, optimizing protein sequences for improved stability and efficiency, and engineering proteins for therapeutic or industrial applications.

For protein structure and function, the agent predicts folding patterns and simulates structural changes in response to mutations or environmental conditions. In protein-protein interaction analysis, it predicts how proteins interact to form complexes and assesses the implications of these interactions on cellular processes. The agent also identifies potential PTM sites and simulates the effects of these modifications on protein stability and function. In protein functional annotation, it annotates proteins with functional domains, classifies them into families, and predicts enzyme activity, binding affinity, and the effects of genetic mutations on protein function. In protein design and engineering, the agent designs proteins with specific functions, optimizes sequences for better stability and catalytic efficiency, and engineers proteins to target specific biological processes. These capabilities are crucial for understanding protein biology and have significant applications in drug discovery, disease modeling, and synthetic biology, and all of these biological tasks can be achieved by LLMs [64].

### 4.2. Organelle AI Agent

Organelle AI agents are advanced platforms designed to integrate multi-modal data and AI technologies to explore the diversity, biological functions, and dynamic behaviors of various cellular organelles. These organelles are specialized subcellular structures that perform vital functions critical to cellular health and organismal homeostasis. The most well-known organelles include the nucleus, mitochondria, endoplasmic reticulum, Golgi apparatus, lysosomes, peroxisomes, and centrosomes, among others. Each of these plays distinct roles in cellular processes, such as protein synthesis, energy production, waste management, and cell division. Within this framework, the Mitochondrial AI Agent is specifically designed to explore the diversity, biological functions, and dynamic behaviors of mitochondria in both healthy and diseased conditions. Mitochondria are essential organelles responsible for ATP production via oxidative phosphorylation (OXPHOS), thereby fulfilling the cellular energy requirement [65]. Beyond energy production, mitochondria are central to the regulation of apoptosis, signal transduction, immune responses, and redox balance. Their unique circular mitochondrial DNA (mtDNA), characterized by a high mutation rate, has been strongly associated with a variety of diseases, including neurodegenerative disorders [66], cancers [67, 68], metabolic syndromes [69], and immune-related diseases [70]. Recent research has expanded the focus of mitochondrial studies beyond their traditional functions to include their dynamic behaviors, intercellular transfer, and molecular characteristics, revealing the multifaceted roles mitochondria play in cellular biology [71]. **Figure 5** shows the predefined detachable biological tasks of the Mitochondrial AI Agent.

The Mitochondrial AI Agent addresses several research domains within the study of mitochondria, including their morphology, distribution, lineage tracing, heterogeneity, and transfer. Within these domains, the AI Agent performs a wide range of tasks to deepen our understanding of mitochondrial functions and behaviors. It predicts the presence of mitochondrial DNA variants and assesses their potential pathogenicity, offering insights into genetic factors associated with disease. The agent also predicts mitochondrial protein structures and their interactions, contributing to the understanding of mitochondrial mechanisms at the molecular level. Additionally, the Mitochondrial AI Agent models key mitochondrial activities, including energy metabolism, membrane potential, neuronal activity, and apoptosis pathways, simulating how mitochondria contribute to cellular processes. It also predicts core mitochondrial functions, including ATP generation efficiency, calcium ion regulation, reactive oxygen species (ROS) control, and other dynamic processes by integrating multi-omics data. Furthermore, the agent studies mitochondrial behaviors such as fusion and fission dynamics, regulation of nuclear gene expression, and inter-organelle communication, all of which are crucial for maintaining cellular health and homeostasis.

To accomplish these tasks, the Mitochondrial AI Agent leverages cutting-edge AI tools tailored to specific aspects of mitochondrial functions. For example, tools like DeepMito [72] are used for advanced microscopy image analysis, while MitoMap [73] supports mtDNA sequencing and pathogenicity scoring. MitoTrace [74] is used to track mitochondrial variants. MitoRelID (Mitochondrial Related Interaction Descriptor) facilitates the assessment of mitochondrial drug-target interactions, enabling personalized drug discovery and treatment strategies. By integrating these tools, the Mitochondrial AI Agent generates high-resolution annotations of mitochondrial disease characteristics, constructs dynamic models of mitochondrial transfer networks, predicts potential biomarkers for diseases, and identifies therapeutic targets. The ultimate goal of the Mitochondrial AI Agent is to significantly enhance our understanding of mitochondrial dysfunction at molecular, cellular, and systemic levels. For instance, by evaluating mitochondrial ATP generation efficiency and energy metabolism, the agent can uncover how cancer cells adapt to hypoxic conditions. Similarly, by predicting mitochondrial transfer behaviors, it can explore strategies to prevent the propagation of damaged mitochondria or promote the transfer of healthy mitochondria, with potential implications for improving therapeutic outcomes.

### 4.3. Cell AI Agent

Cells are the fundamental building blocks of all living organisms, representing the smallest units of life capable of independently carrying out the essential processes required for an organism's survival. At the core of cellular function, the interactions among DNA, RNA, and proteins enable cells to perform a broad range of vital tasks necessary for life. DNA stores genetic information, RNA transfers this information, and proteins execute the diverse functions required to maintain cellular operations. Core cellular functions encompass energy production and metabolism, signal reception and transduction, protein synthesis and secretion, cell division and growth, DNA repair and maintenance, immune defense, transport and communication, cellular homeostasis, and mechanical movement and contraction. Advancements in high-throughput sequencing technologies have significantly enhanced our ability to examine and understand the molecular mechanisms underlying these fundamental cellular functions. These technologies enable the analysis of gene expression, protein function, metabolite profiles, and genomic integrity, providing powerful tools for studying cellular processes in both healthy and diseased conditions. The Cell AI Agent is designed to integrate multi-modal data, including genomics, proteomics, morphomics, single-cell multi-omics, and spatial omics, with deep learning algorithms to simulate and predict cellular functions, behaviors, molecular features, heterogeneities, and interactions. By incorporating clinical data, this agent bridges the gap between basic cellular research and clinical applications, facilitating disease diagnosis and treatment.

The Cell AI Agent operates across multiple core domains, beginning with the construction of intracellular metabolic networks to model biochemical pathways. It predicts cellular responses to environmental changes, such as hypoxia or toxin exposure, and links molecular dysfunctions to clinical phenotypes and disease associations. In exploring cellular behaviors, the agent models processes, including cell dissemination patterns, cycle progression, differentiation trajectories, and the dynamics of apoptosis and autophagy. It also provides insights into the spatial distribution and composition of cells within tissues, contributing to tissue modeling and developmental studies. On the molecular level, the agent predicts gene expression profiles and regulatory networks, detects both genetic and epigenetic variations, and evaluates the pathogenicity of mutations. This functionality is crucial for precision medicine and the identification of drug targets. The agent also excels in analyzing cellular heterogeneity by classifying cell subtypes, predicting their functions, mapping gene expression throughout the cell life cycle, and identifying functional subtypes within populations. This is critical for understanding complex processes such as cancer evolution and immune responses. A key strength of the Cell AI Agent lies in modeling cellular interactions. It simulates cell-cell signaling, cell-matrix interactions, and molecular changes during intercellular communication. By predicting ligand-receptor interactions, the agent helps elucidate mechanisms involved in immune regulation, tissue repair, and tumor microenvironments. Clinically, the Cell AI agent contributes by identifying disease biomarkers, predicting drug responses at the cellular level, and optimizing therapeutic molecules, including antibodies and gene editors. It also supports the design of cell-based therapies such as CAR-T cells, predicts disease risks, and assesses immune evasion strategies in cancer. **Figure 6** shows the predefined detachable biological tasks assigned to the Cell AI agent.

The integration of single-cell multi-omics data is one key strength of the Cell AI Agent. By combining gene expression, chromatin accessibility, DNA methylation, and protein activity, the agent offers a comprehensive view of cellular states and functions. This multi-omics approach helps overcome challenges such as technical noise and batch effects, enabling more robust analysis and interpretation of cellular biology. Through the integration of multi-scale data and the application of advanced computational methods, the Cell AI Agent significantly enhances our understanding of cellular processes and fosters innovations in diagnostics, personalized treatment, and regenerative medicine.

To support these applications, the Cell AI Agent employs advanced computational algorithms designed for tasks such as cell clustering, cell type annotation, cell cycle modeling, and the analysis of cell-cell interactions. Clustering methods, including graph-based and density-based approaches, enable the identification of cell groups with similar gene expression profiles. Among these, the Louvain and Leiden algorithms are particularly effective for clustering in scRNA-seq analysis. Following clustering, cell type annotation is performed by assigning biological identities to each cluster based on known marker genes. Reference databases like CellMarker [75] and PanglaoDB [76] provide essential resources for both manual and automated annotation. Furthermore, tools like Monocle[77] and STREAM [78] allow researchers to reconstruct developmental trajectories and visualize gene expression patterns along pseudotime, offering valuable insights into cellular transitions and lineage relationships. To enhance these capabilities, the cell-level AI Agent further incorporates advanced AI-based tools specifically designed to overcome major challenges in single-cell analysis. For example, to address data sparsity in scRNA-seq, generative models such as Generative Adversarial Networks (GANs) [79] have been applied for imputation. To better capture dynamic processes such as the cell cycle, self-supervised learning frameworks [80] [81] have been proposed to model its circular nature and represent pseudotime along this trajectory. Similarly, novel extensions of RNA velocity, such as the lazy probability model [82], provide the ability to estimate not only the directionality but also the probability of cell state transitions. Beyond transcriptomics, transformer-based large language models (LLMs) have recently been applied to predict drug sensitivity [83], highlighting the expanding role of cell-level AI Agent in integrating diverse data modalities.

## 4.4. Tissue AI Agent

The Tissue AI Agent is an advanced computational platform designed to study tissue-level biological processes by integrating multi-omics data and spatial biology. Tissues are composed of groups of similar cells that work together to perform specific functions, often embedded within an extracellular matrix (ECM) that provides structural support and biochemical signaling necessary for tissue integrity and function. At the tissue level, the interactions between cells, the ECM, and molecular signals are crucial for maintaining homeostasis and responding to stress or disease. Understanding tissue structure and function requires the analysis of cellular interactions, molecular signaling pathways, and tissue-level dynamics. The Tissue AI Agent bridges the gap between molecular, cellular, and organ-level studies by leveraging AI-driven approaches. The Tissue AI Agent performs several key tasks, including modeling and simulation of tissue dynamics, discovery of novel biomarkers and regulatory networks, construction of comprehensive tissue knowledge bases, and prediction of patient-specific tissue responses to treatments. In tissue dynamics, the agent integrates spatial omics data and agent-based modeling to simulate tissue environments. This enables a deeper understanding of how tissues develop, maintain homeostasis, and respond to perturbations such as injury or disease. For example, the agent can model tissue repair following injury or track the progression of diseases like fibrosis and cancer.

Through these simulations, the Tissue AI Agent predicts emergent properties such as tissue differentiation, regenerative potential, and disease-related changes in tissue architecture. To advance tissue biology research, the agent employs self-supervised learning techniques to uncover tissue-specific regulatory networks, rare cell populations, and novel biomarkers. This autonomous discovery capability enables deeper insights into disease mechanisms and supports the development of diagnostic strategies. The Tissue AI Agent integrates multi-omics data-such as single-cell transcriptomics, proteomics, and epigenomics, with spatial mapping to construct a comprehensive tissue knowledge base. By using AI-driven knowledge graphs, it uncovers functional tissue interactions and sheds light on both healthy and diseased states. In the context of personalized medicine, the Tissue AI Agent predicts patient-specific responses to treatments by analyzing tissue heterogeneity. It can detect early disease signatures, such as those present in cancer microenvironments or neurodegenerative disorders, thereby facilitating the development of precision therapies. Furthermore, the agent contributes to biofabrication and synthetic tissue engineering by simulating stem cell differentiation and optimizing conditions for tissue regeneration. **Figure 7** shows the Tissue AI-Agent predefined detachable biological tasks performed by the Tissue AI Agent.

The Tissue AI-Agent employs several advanced AI tools and algorithms to achieve these tasks. Including: (1) Deep learning models, such as convolutional neural networks (CNNs) and Vision Transformers (ViTs), are used to analyze histopathological features and spatial omics data. SEQUOIA [84] and RNAPath [85] are transformer-based models that leverage whole slide histology images to respectively predict cancer transcriptomic profiles and spatially localize RNA expression, enabling cost-effective genomic analysis and revealing the spatial interplay between tissue morphology and gene expression. (2) Graph neural networks (GNNs) are utilized to model cell-cell interactions and tissue architecture, while reinforcement learning (RL) simulates tissue responses to external perturbations. STCase [86] and Ceograph [87] are graph neural network-based models that analyze spatial transcriptomics and pathology images, respectively, to uncover niche-specific cell-cell communication and spatial cell organization, enabling fine-grained mapping of microenvironmental interactions and accurate prediction of clinically relevant tissue features. And, RL-GenRisk [88] uses reinforcement learning to simulate tissue responses to pathological perturbations, identifying ccRCC (clear cell renal cell carcinoma) risk genes through a graph-based Markov decision process. (3) SSL techniques allow the agent to discover hidden spatial patterns within tissue samples without requiring labeled data, enabling deeper insights into tissue organization and behavior. BIDCell [89] and Kasumi [90] are self-supervised learning models that uncover spatial patterns in tissue without labeled data, BIDCell links gene expression to cell morphology for accurate segmentation, while Kasumi identifies persistent spatial neighborhoods associated with disease outcomes. By integrating these computational techniques, the Tissue AI-Agent is capable of predicting how tissues evolve in response to physiological and pathological stimuli. This ability to model tissue dynamics provides new opportunities for understanding disease progression, tissue regeneration, and therapeutic interventions, and the development of targeted therapeutic interventions.

### 4.5. Organ AI Agent

The Organ AI Agent is designed to model and analyze the biological functions of individual organs by integrating multi-omics data, clinical information, and imaging. This integration provides comprehensive insights into organ-specific diseases, functions, and responses to stimuli. The agent constructs detailed anatomical and functional models of organs, simulates their behavior in various conditions, and predicts responses to environmental changes or treatments. The Organ AI Agent focuses on several core tasks, including modeling organ function, analyzing organ pathophysiology, identifying molecular features specific to the organ, examining heterogeneity within the organ, and studying tissue interactions. Each task is based on data from multiple sources, such as genomics, transcriptomics, proteomics, physiological measurements, and imaging, which are collectively integrated to generate a holistic organ profile. **Figure 8** shows the predefined detachable biological tasks designed for the Organ AI Agent.

Organ function modeling involves simulating and predicting an organ's functional status under both healthy and diseased conditions. This includes evaluating the impact of environmental factors on organ function, such as nutrient availability, metabolic states, or disease progression. By combining anatomical imaging (e.g., MRI or CT) with molecular data, the Organ AI Agent can simulate dynamic changes in organ function over time, including responses to injury, disease, or therapeutic interventions. In organ pathophysiology analysis, the agent examines the dynamic alterations that occur in response to specific diseases. It predicts the underlying causes of organ dysfunction, providing insights into disease mechanisms and helping to identify potential therapeutic targets. For example, the agent may simulate the progression of fibrosis following chronic injury or model tumorigenesis driven by specific genetic mutations. Additionally, the Organ AI Agent also identifies molecular features associated with normal organ function and pathological states, such as particular marker genes or molecular pathways. By analyzing transcriptomics and proteomics data, the Organ AI agent can identify molecular characteristics linked to disease progression and predict how these features evolve across different disease states. Another key function of the Organ AI Agent is examining organ heterogeneity. It predicts spatial and functional variability within the organ, identifying regions with distinct molecular and cellular profiles. This analysis can be used to study tissue-specific changes, such as how different cell types, metabolic activities, or molecular distributions contribute to organ function and dysfunction. Additionally, the agent models tissue interactions within the organ. It simulates the flow of signals and materials between different tissue regions, such as the interaction of immune cells with tissues during inflammation or the influence of signaling pathways during tissue repair. By integrating spatial omics and imaging data, the Organ AI Agent provides a detailed view of how various regions of the organ function coordinate functionally and how they respond to external stimuli. Finally, the Organ AI Agent supports drug response and toxicity analysis by predicting the impacts of treatments on specific organs. It can optimize organ engineering and regenerative medicine strategies, such as organ transplantation, by simulating the adaptation process of transplanted organs in the host and predicting the outcomes of various therapeutic approaches.

The Organ AI Agent also leverages several advanced technologies and methodologies to model and analyze the dynamic interactions between organs. Organ-on-a-chip technology employs microfluidic chips to simulate in vitro organ environments and inter-organ interactions. It is ideal for dynamic physiological study, complex disease modeling, and drug evaluation.

Organoid culture involves the generation of miniature, three-dimensional organ structures from primary tissues or stem cells under controlled conditions[91]. These models offer valuable insights into organ development, disease processes, and personalized medicine. Finite element analysis (FEA) is used to simulate and analyze the biomechanical properties of tissues, such as muscle behavior, under different conditions [92]. It can facilitate the understanding of how mechanical factors influence organ function. Virtual human modeling and simulation create detailed three-dimensional models of the human body [93], allowing for the simulation of organ movement and interactions. This is particularly useful for medical device development, surgical planning, and the design of individualized treatments. By integrating these diverse methods, the Organ AI-Agent delivers a comprehensive understanding of organ function, interactions, and responses to external factors.

Some studies have integrated imaging and genetic analysis to investigate organ function and its association with diseases, and the Organ AI-Agent will incorporate these capabilities [94, 95]. It begins with MRI and CT imaging to identify functional regions of interest (ROIs) within the organ, providing high-resolution anatomical and functional data. Once these ROIs are defined, the agent uses genome-wide association studies (GWAS) to link genetic variants to specific features within these regions, thereby identifying genetic loci that influence organ function or pathology. To infer causal relationships, Mendelian randomization (MR) is employed, using genetic variants as instrumental variables to determine if genetic factors directly affect organ function, independent of confounding variables. By integrating imaging with GWAS and Mendelian randomization, the agent constructs a comprehensive model of organ behavior, linking structural features with genetic pathways to identify potential therapeutic targets for precision medicine.

### 4.6. Organ System AI Agent

The Organ System AI Agent is an advanced platform designed to model and analyze inter-organ communication, homeostasis, and responses to perturbations across multiple organs. By integrating a vast array of biological, physiological, environmental, and molecular data through sophisticated AI models, the agent provides a comprehensive understanding of organ-to-organ interaction within the human body. Organ systems are fundamental for maintaining physiological functions, and their coordinated communication, both local and distant, is critical for adapting to diseases and ensuring overall homeostasis. Recent studies highlight the interconnection between organ systems in conditions such as cardiovascular, immune, and neurodegenerative diseases, revealing how dysfunction in one organ can influence the function of others. A notable example is the gut-brain axis, where gut-derived factors like α-Syn have been implicated in the early diagnosis of Parkinson's disease [96]. By capturing these complex inter-organ relationships, the Organ System AI Agent enables disease prediction, biomarker discovery, and optimization of therapeutic strategies.

The Organ System AI Agent integrates multiple datasets, including multi-omics (genomics, metabolomics, metagenomics), physiological data (e.g., ECG, blood glucose, sleep rhythm), anatomical data (e.g., liver thickness, left ventricular volume), whole-body imaging data (CT, MRI, ultrasonography), and environmental factors (e.g., smoking, drug use) to carry out a wide range of downstream biological tasks. Image data provides detailed anatomical structures, while non-image data offers insights into organ functions and overall health conditions. The core tasks of the Organ System AI-Agent include analyzing the coordination of multi-organ functions, modeling dysfunction in systemic diseases across organs, simulating multi-organ pathological cascades, and predicting drug responses. The agent also simulates cross-organ compensation and repair processes following injury, constructs multi-organ metabolic networks, and infers biomechanical interactions between organs to predict functional and phenotypic changes. **Figure 9** shows the predefined detachable biological tasks designed for the Organ System AI Agent.

To achieve these tasks, the Organ System AI-Agent relies on advanced AI tools such as nnU-Net for biomedical image segmentation [97], TMO-Net for capturing self-modal and cross-modal features [98], and BioBERT [99] for biomedical text mining. Additionally, Mendelian randomization is employed to identify causal relationships between organs based on genomic variants, enabling the investigation of how genetic factors influence inter-organ interactions and the development of systemic diseases. By integrating this approach with other analytical tools, the Organ System AI Agent captures the anatomical, molecular, and environmental features of organs under various biological states.

### 4.7. Body system AI Agent

As illustrated in **Figure 10**, the Body System AI-Agent encompasses several core functions that operate both within and across organ systems, including: (1) Intra-System Functional Coordination: The Body System AI-Agent models the dynamic interplay between organs within a single physiological system. For instance, it analyzes how the heart, lungs, and kidneys coordinate to regulate circulatory and respiratory functions. By integrating imaging data (MRI, CT), electronic medical records (EMR), and multi-omics datasets (genomics, metabolomics), the agent simulates how perturbations in one organ influence others, providing insights into system-level homeostasis and organ resilience. (2) Inter-System Functional Coupling: Beyond individual systems, the agent captures crosstalk between different organ systems, such as the nervous, immune, and endocrine systems. It simulates how neural signaling affects musculoskeletal function or how hormonal cascades, from the hypothalamus to adrenal glands, regulate physiological responses. This modeling enables discovery of regulatory axes and feedback loops that underlie whole-body function.

(3) Systemic Disease Propagation: The Body System AI-Agent can predict how diseases spread across organ systems. For example, it models how chronic inflammation originating in the gut may trigger immune responses or dysfunctions in distant organs, contributing to systemic disorders like autoimmune diseases or multi-organ failure. These simulations help uncover disease pathways and inform early intervention strategies. (4) Cross-Organ Disease Dynamics: The agent simulates disease progression across organ systems, such as how cardiovascular disease may impair renal function or how diabetes can affect neural and ocular tissues. By modeling disease trajectories and interaction networks, the agent aids in identifying critical nodes for therapeutic targeting and in understanding systemic comorbidities. (5) System-Level Metabolic Dynamics: A key strength of the Body System AI-Agent lies in modeling metabolic flow between and within organ systems. It simulates how disruptions in lipid or glucose metabolism affect organs such as the liver, pancreas, and cardiovascular system. These insights are vital for studying metabolic syndromes and their systemic consequences, including obesity, type 2 diabetes, and fatty liver disease. (6) Biomechanical and Structural Interactions: The agent integrates mechanical forces and structural biology to simulate how physical stressors affect organ behavior. For instance, it can predict the heart's adaptation to pressure overload or how skeletal muscle responds to repetitive motion. This functionality supports modeling of injury, aging, and adaptive responses under different physical conditions. (7) Personalized Therapy Optimization: The Body System AI-Agent plays a central role in precision medicine by simulating individualized responses to drugs and interventions across multiple organ systems. By leveraging EMR, clinical trial data, and omics profiles, it predicts therapeutic efficacy and toxicity, optimizes combination treatments, and tailors drug regimens to maximize systemic benefit while minimizing adverse effects.

In addition, EMR data plays a crucial role in understanding human systems biology, as it captures the macroscopic manifestations of dynamic changes occurring across multiple biological levels. While numerous studies have been conducted to analyze the physiological changes of different organs or systems based on EMR data, these studies face a significant limitation: they cannot simultaneously capture all organ-level data from the same individual. This constraint hinders holistic insights into the body's systemic complexity. To address this, we propose incorporating LLMs into the Body System AI-Agent framework. By mining EMR data, LLMs can identify physiological signals associated with specific organs, shared across multiple organs, or linked to comorbid disease states. This capability enables the construction of a high-dimensional, integrated network that maps physiological signals to organs and organ systems. Through such integration, the Body System AI-Agent can reveal the interdependencies among biological systems, offering deeper insights into human health and disease and enabling more precise, system-aware therapeutic strategies.

## 5. Applications of the Full-body AI-Agent System: Case Studies

The Full-body AI-Agent framework enables a systematic, multi-scale approach to dissecting disease mechanisms and optimizing therapeutic interventions. Here, we present two case studies: the exploration of Lung cancer metastasis (**Figure 11**) and Drug development (**Figure 12**), demonstrating the framework's potential in both disease research and clinical translation.

**Case 1: A three-phase metastasis scoring framework leveraging full-body AI-Agents**

Metastasis, the spread of malignant cells from a primary tumor to distant organs, is responsible for the majority of cancer-related deaths. It is widely conceptualized as a series of sequential events, termed the invasion-metastasis cascade [100]. Tumor metastasis unfolds in three successive phases (**Figure 11A**). **Figure 11B** shows the occurrence and progression of tumor metastasis. In the Initiation phase, epithelial tumor cells undergo epithelial-to-mesenchymal transition (EMT) to gain motility and invasive potential [101]; then, in coordination with cancer stem-cell traits [102], metabolic reprogramming [103], angiogenic remodeling [104], immune evasion [105] and clonal selection [106] to breach the basement membrane and intravasate. During the Dissemination phase, circulating tumor cells (CTCs) face multiple survival challenges, including hemodynamic shear stress, anoikis, and immune attack. To overcome these barriers, CTCs employ mechanisms such as platelet cloaking and endothelial adhesion, allowing them to traverse the vasculature [107]. Finally, in the Colonization phase, disseminated tumor cells extravasate into distant tissues and often revert to an epithelial phenotype through mesenchymal-to-epithelial transition (MET) [108]. These cells may enter a dormant state and, under the influence of late-stage immune suppression and supportive microenvironmental cues, become reactivated to form overt macroscopic metastases. Importantly, tumor colonization at distant sites is governed by organ-specific microenvironmental signals that uniquely shape the trajectory of metastatic outgrowth [109]. For instance, in the lung, circulating tumor cells are entrapped and activated by neutrophil extracellular traps (NETs) and endothelial CXCL12 gradients, which cooperate to awaken dormant cells and promote early micrometastatic seeding [110]. In the liver, activated hepatic stellate cells impose a fibrotic niche that suppresses natural killer cell-mediated clearance, enforcing prolonged dormancy of disseminated tumor cells until niche remodeling permits their reawakening [111]. In the brain, astrocyte-derived exosomal microRNAs induce PTEN downregulation in tumor cells, creating a growth-permissive niche that primes metastatic cells for proliferation [112]. These examples illustrate how each organ's unique stromal and immune landscapes dictate the survival, dormancy, and eventual outgrowth of disseminated cancer cells. The complexity and organ-specific nature of metastasis underscore the need for a global perspective that integrates multimodal data and mechanistic insights across molecular, cellular, tissue, organ, and systemic scales to fully elucidate its dynamic progression and interdependencies.

Integrating across all biological scales is essential to understanding tumor metastasis. Despite significant advances at individual biological scales, from gene discovery to organ-level imaging, a comprehensive understanding of how molecular events cascade through cells, tissues, and organs to manifest as systemic diseases remains elusive. To bridge this gap, we introduce the Full-Body AI-Agent, a unified framework composed of seven foundational AI modules, each corresponding to a specific level of biological organization. Based on multi-omics and multi-modal data, and by integrating insights from molecules, organelles, cells, tissues, organs, inter-organ interactions, and full-body physiology, this framework enables the reconstruction of complex biological networks with unprecedented resolution and interpretability. Any AI-driven analysis that integrates data across two or more of these levels is considered within the scope of the Full-Body AI Agent. The Metastasis AI Agent serves as a prototypical example to model and decode the multi-step, multi-organ nature of cancer metastasis.

We anchored our Metastasis AI Agent on three distinct yet interdependent metrics: the Initiation score, Dissemination score, and Colonization score (**Figure 11C**). These scores correspond to the fundamental biological challenges that tumor cells must overcome during their progression from the primary site to distant metastatic sites. The Initiation score assesses a tumor's ability to detach from the primary site, invade surrounding tissues, and evade local immune surveillance. The Dissemination score quantifies the capacity of CTCs to survive hemodynamic stress, resist immune attack, and adhere to the endothelium at distant sites. The Colonization score evaluates the likelihood that disseminated tumor cells will enter a dormant state, adapt to a new niche, and ultimately resume proliferation to form overt metastases. By focusing on these three scores, the Metastasis AI Agent pinpoints the specific vulnerabilities at each stage, guides targeted data integration across biological scales, and enables phase-specific risk assessment and therapeutic intervention. Here, we detail how Metastasis AI Agent achieves a Full-Body understanding of Metastasis through the basic AI Agent at seven biological levels and cross-scale quantification of metastatic potential via three scores (**Figure 11D**).

To evaluate the Initiation core, the Molecule AI Agent first quantifies the expression of EMT drivers (such as SNAIL and TWIST), stemness regulators (including NANOG and SOX2), and genetic alterations that confer a selective advantage to tumor cells. Next, the Cell AI Agent analyzes single-cell phenotypes to gauge the prevalence of hybrid EMT states, cancer stem cell-like clusters, and shifts in metabolic pathway activity indicative of glycolytic reprogramming versus oxidative phosphorylation. Concurrently, the Tissue AI Agent evaluates angiogenic remodeling by quantifying microvessel density and perivascular cell infiltration from segmented immunohistochemistry data. The Organ AI Agent extracts metrics of perfusion heterogeneity and vessel permeability metrics from multiparametric imaging. In parallel, the Cell and Tissue AI-Agents collaboratively score local immune escape by identifying PD-L1 expression patterns and the presence of immunosuppressive cell populations. Combined through a multilayer ensemble, these heterogeneous features yield a continuous Initiation Score that robustly differentiates tumors based on their early invasive potential.

The Dissemination Score assesses the ability of circulating tumor cells to survive hemodynamic shear stress, resist anoikis, and adhere within distant vasculature. The Cell AI Agent profiles CTCs to identify transcriptional programs that promote anchorage-independent survival, including anoikis resistance and upregulation of platelet-interacting receptors. Meanwhile, the Organ-System AI Agent constructs patient-specific hemodynamic simulations from four-dimensional flow MRI and computational fluid dynamics, pinpointing vascular regions of low shear stress that favor tumor cell arrest and platelet cloaking. The Tissue AI Agent quantifies endothelial adhesion molecule expression, such as ICAM-1 and VCAM-1, through spatial transcriptomics data to map vascular adhesion hotspots. By integrating intrinsic survival signatures with organ-axis fluid dynamics and adhesion landscapes, the Dissemination Score captures a cell's capacity to transit, persist, and arrest within the circulatory system.

To assess the Colonization Score, the Molecule AI-Agent examines single-cell metabolic profiles of dormant cells to detect shifts toward fatty-acid oxidation and enhanced antioxidant pathways. The Cell AI-Agent assesses the extent of MET by measuring the re-expression of epithelial markers and the restoration of junctional complexes. The Tissue AI-Agent characterizes the extracellular matrix by profiling its composition, stiffness gradients, and vascular density through multiplexed imaging to define niche permissiveness. The Organ AI-Agent integrates PET tracer uptake and metabolomics data to map organ-specific nutrient availability, while the Organ-System AI-Agent models cross-organ immune-suppression axes, such as the recruitment of bone marrow-derived myeloid cells, to capture late-stage immune editing. By harmonizing these multi-scale features, the Colonization Score predicts the likelihood of dormant cell survival, niche integration, and metastatic outgrowth.

To provide a concrete example of how metastatic potential is quantified during the Initiation phase, we analyzed single-nucleus RNA sequencing (snRNA-seq) data from a primary non-small cell lung cancer (NSCLC) tumor (sample N2254 [113]). The Initiation Score integrates molecular, cellular, tissue, and organ-level features, each extracted by the corresponding AI Agent using specialized computational tools.

At the molecular level, the Molecular AI Agent incorporates epithelial-to-mesenchymal transition (EMT) drivers, cancer stemness regulators, metabolic reprogramming signatures, and genetic alterations conferring selective growth advantages. snRNA-seq preprocessing was performed using Scanpy, including quality control, normalization, log-transformation, and selection of highly variable genes. Dimensionality reduction and clustering distinguished malignant from non-malignant epithelial cells, with downstream analyses focused on the malignant subpopulation. Marker gene sets representing epithelial and mesenchymal states, EMT drivers, stemness regulators, glycolysis, and oxidative phosphorylation (OXPHOS) were curated. Using Scanpy's score_genes function, per-cell scores were calculated.

A hybrid EMT score was defined as the minimum between normalized epithelial and mesenchymal scores, with the top 20% of cells classified as hybrid EMT. A metabolic reprogramming score was calculated as the glycolysis-to-OXPHOS ratio. All features were normalized to a 0-1 scale. A comprehensive Molecular Initiation Score was then derived using principal component analysis (PCA), with the first principal component providing weights for feature integration. The mean Molecular Initiation Score across malignant cells was 0.257 (range: 0.014-0.753). Cells in the top 25% (threshold = 0.324) were classified as high initiation potential, yielding 1,625 cells. Cluster-level aggregation identified Cluster 6 as having the highest mean score (0.326). Differential expression analysis revealed significant upregulation of CD44, FN1, and VIM, consistent with invasive phenotypes. Pathway enrichment confirmed activation of EMT transcription factors and stemness regulators in high-initiation cells.

At the cellular scale, the Cell AI Agent evaluated the prevalence of hybrid EMT phenotypes and cancer stem cell-like clusters. Clustering of malignant cells using Louvain clustering enabled mapping of cell states against hybrid EMT and stemness scores [114]. This revealed distinct cellular subpopulations enriched for high initiation potential, highlighting the contribution of cellular heterogeneity to metastatic initiation.

Although the Tissue AI Agent and Organ AI Agent were not directly applied to this dataset, their roles are critical in expanding the Initiation Score quantification. The Tissue AI Agent extracts angiogenic and immune features from histopathology or spatial omics data, such as microvessel density, angiogenic remodeling, and infiltration by immunosuppressive cell types. The Organ AI Agent quantifies perfusion heterogeneity and vascular permeability from multiparametric imaging. Together, these higher-level metrics integrate with molecular and cellular features to refine Initiation Score estimation and capture tumor physiology across scales.

The Initiation Score is ultimately computed through a multilayer ensemble framework that integrates outputs from all AI Agent levels. In this case, the Molecular and Cell AI Agents provided the primary input, yielding a robust stratification of cancer cells and clusters by early invasive potential. This demonstration illustrates how multi-scale computational integration empowers the Metastasis AI Agent to quantify and predict metastatic initiation with high precision.

**Case 2: The application of the Full-Body AI agent platform in drug development**

The enduring human aspiration for health and longevity continues to drive scientific efforts in drug development. Yet, the complexity of the human body remains only partially understood, presenting significant obstacles to the design, evaluation, and deployment of effective therapies. In **Figure 12A**, drug development is conceptualized as a biological staircase, one that must be traversed across multiple interconnected levels of biological organization. This multi-scale perspective highlights the importance of integrating molecular mechanisms with systemic physiological responses. It illustrates how each layer, from protein structures to whole-body dynamics, contributes to the success or failure of a therapeutic strategy. At the whole-body system level (①), drug effects manifest as changes in global neuroendocrine, immune, and behavioral states. These systemic responses are shaped by the integrated dynamics of multiple organ systems and reflect the ultimate therapeutic outcome, which can be either an improvement or an adverse reaction. This level defines the clinical endpoints by which treatment efficacy is judged, including survival, symptom relief, or improvement in quality of life, thus anchoring the top of the biological staircase. Descending to the organ system level (②), interorgan communication becomes central. A notable example is the heart-brain axis, where physiological signals such as cardiac rhythm and cerebrovascular pressure allow the heart to influence brain function[115]. Such interactions underscore the importance of accounting for multi-organ crosstalk during drug development. At the organ level (③), anatomical structures become the focus point. Here, the brain is shown with a localized tumor, representing a pathological region embedded within a complex functional organ. The structural localization of disease introduces challenges in spatial drug delivery and highlights the need for targeting strategies that consider both regional anatomy and functional compartmentalization. The tissue and microenvironment level (④) provides a more granular perspective, focusing on the tumor microenvironment, including the tumor surrounding matrix, immune context, and vasculature, notably the blood-brain barrier (BBB). The BBB represents a formidable obstacle to drug permeability, representing both a physical and biological barrier that must be overcome to achieve adequate drug concentrations at the site of action. At the cellular level (⑤), the tumor is no longer viewed as a bulk mass but rather as a heterogeneous population of individual cells, each exhibiting distinct transcriptional profiles, phenotypic behaviors, and drug sensitivities. Achieving single-cell resolution is essential for uncovering drug resistance, clonal evolution, and cell type-specific responses that are often masked in bulk analyses.

Descending further, the subcellular level (⑥) focuses on intracellular alterations such as organelle dysfunction, disrupted signaling pathways, and metabolic stress, which collectively shape how a cell perceives and responds to a therapeutic agent. This level captures early mechanistic signals that link molecular intervention to phenotypic change, offering critical insights into both drug efficacy and toxicity. At the base of the staircase lies the molecular level (⑦), where drug-target interactions are determined by atomic-scale features, such as protein conformation, binding pocket geometry, and ligand specificity. It is at this foundational level that pharmacological action originates and where medicinal chemistry plays a central role in optimizing structure-activity relationships to initiate therapeutic cascades.

Drug development strategies have expanded significantly in recent years, yet clinical translation remains a persistent bottleneck. In practice, more than 90% of drug candidates deemed "safe and effective" in animal models ultimately fail in human clinical trials, primarily due to insufficient efficacy or unacceptable toxicity [116, 117]. This high attrition highlights the inherent limitations of animal models in predicting human outcomes. Recognizing this, the U.S. Food and Drug Administration (FDA), through its Reducing Animal Testing Roadmap, has identified interspecies immunogenicity and physiological divergence as the key contributors to this disconnection. As a result, the FDA has formally endorsed the development of human-relevant platforms such as organoids to reduce reliance on animal studies and enhance translational success.

Organoids and organ-on-a-chip systems, which combine three-dimensional human cell self-assembly with microfluidic biomimicry, offer high-fidelity recapitulation of localized organ function [118]. A notable milestone was achieved when a lung-on-chip model enabled Azeliragon to advance directly to Phase II clinical trials, demonstrating the potential of such systems to accelerate the transition from in vitro discovery to human validation [119]. However, as these technologies advance, their limitations must also be critically examined from a systems-level perspective. Two fundamental constraints are particularly salient. First is the boundedness of biological systems: the human body is a closed, dynamically coupled network of interacting organs. Most current chip models focus on single organs or limited pairings, making it difficult to recapitulate the cascading metabolism and distal toxicities that unfold along the liver-kidney-gut-immune axis. Second is the orderliness of biological processes: many chronic toxicities emerge only after weeks or months of accumulation and systemic feedback. Yet, the duration and stability of current chip cultures rarely support such long-term observation windows. Recent reviews have also noted structural limitations. For example, even advanced liver organoids often lack critical cell populations such as cholangiocytes and vascular endothelial cells, which impairs their ability to model drug bioactivation and metabolic toxicity with sufficient accuracy [120, 121]. New strategies for preclinical screening are urgently needed to better predict clinical outcomes, thereby expediting drug development while maintaining cost-effectiveness.

To systematically address the limitations of conventional preclinical models and enhance the fidelity of translational research, we introduce the Drug AI-Agent, a core module of the Full-Body AI-Agent framework, as illustrated in Figure 12. Figure 12A outlines the full spectrum of biological hierarchy and its integration into drug development, spanning from molecular structures to systemic physiological responses. Drug development proceeds through six canonical stages (**Figure 12B**), beginning with target identification and compound discovery, progressing through preclinical validation, and culminating in clinical trials, regulatory approval, and post-market surveillance [122]. Within this workflow, organoids and organ-on-chip models are primarily positioned in Step 3, serving as key platforms for preclinical evaluation of safety, efficacy, and pharmacodynamics. **Figure 12C** highlights a broad set of functional requirements that span all stages of drug development, from target validation and immunogenicity screening to pharmacokinetics, toxicity assessment, and post-market failure analysis. Meeting these demands requires scalable, human-relevant, and biologically contextualized modeling across multiple layers of biological organization.

Within the Full-Body AI-Agent framework, the Drug AI-Agent serves as a higher-order integrative module that synthesizes insights from seven foundational AI-Agents to enable comprehensive, system-aware drug development. Each foundational agent leverages level-specific biomedical data to fulfill key tasks. These agents collectively simulate and analyze the whole cascade of drug action. This architecture allows insights generated at the molecular or cellular level to propagate upward through tissues, organs, and entire physiological systems. Conversely, systemic constraints such as inter-organ signaling or immune response feedback downward to inform drug design, delivery, and dosing strategies. In this way, the Drug AI-Agent provides a computationally integrated, causally aware, and physiologically grounded alternative to conventional preclinical workflows, embedding organoid-derived findings into a dynamic whole-body context.

**Figure 12D** shows the biological levels involved in various tasks and biological functions during the drug development process, demonstrating the complexity of multiple biological levels in the drug development process. By aggregating outputs from all these levels, the Drug AI-Agent constructs a dynamic, multi-scale understanding of how molecular interventions propagate through cellular, tissue, and organ systems to shape organism-level responses. This unified view enables more accurate efficacy prediction, systemic toxicity forecasting, and design feedback loops, bridging the gap between in vitro models and in vivo clinical outcomes.

Building on the structural and functional insights described above, the Full Body AI-Agent can execute a comprehensive end-to-end *in-silico* drug discovery pipeline. First, in-house and public compound libraries are virtually screened against predicted binding pockets using high-throughput docking tools such as AutoDock-GPU [123] and GNINA [124]) to generate binding-affinity scores. Promising hits are refined through molecular dynamics simulations (e.g., GROMACS [125]) and free-energy perturbation analyses to optimize binding poses and estimate binding free energy$\Delta G$). The agent then invokes generative models such as DeepChain [126], DiffDock [127], and equivariant diffusion networks to de-novo design analogs optimized for multiple objectives, including potency, selectivity, synthesizability (SA score), and patentability. Each candidate is evaluated through rapid ADMET prediction modules: pkCSM [128] for pharmacokinetics, hERG-Block for cardiac safety, and Tox21 ensemble models for general toxicity profiling. Finally, synthesis routes are proposed via retrosynthetic planning (RETRO models [129], ASKCOS [130]) and prioritized based on cost and synthetic step count. These *in-silico* outputs feed into experimental loops (biochemical assay optimization, SPR/ITC binding validation) that iteratively retrain the agent, effectively closing the design-make-test-analyze cycle.

Recent real-world applications illustrate the transformative potential of AI-agent-driven reasoning in drug development: ISM001-055 (INS018_055), developed by Insilico Medicine, is the first AI-discovered and AI-designed drug candidate for idiopathic pulmonary fibrosis (IPF), identified via generative AI for novel target discovery, now successfully completing Phase IIa trials with dose-dependent lung function improvement, good safety, and FDA orphan drug designation [131]; Halicin, discovered at MIT through AI-based antibiotic screening, represents a novel broad-spectrum antibiotic effective against multiple drug-resistant pathogens, which eradicated otherwise untreatable infections in murine models and exemplifies AI's ability to explore novel chemical spaces despite still being in preclinical stages [132]; Baricitinib, identified by BenevolentAI's platform, was repurposed from rheumatoid arthritis treatment to target SARS-CoV-2–related inflammation, progressing from AI-generated hypothesis to FDA emergency use authorization in just nine months, demonstrating unprecedented acceleration in translational timelines [133]. Collectively, these cases underscore that when multi-agent architectures are applied, especially those with cross-scale integration like the Full-Body AI-Agent, the drug development process can transcend conventional siloed stages. By unifying molecular discovery, organoid modeling, and whole-body simulation within a single reasoning framework, such systems can bridge the translational gap, reduce attrition rates, and enable faster, more cost-effective delivery of safe and effective therapies.

## 6. Current Progress and Challenges

By integrating artificial intelligence across multiple biological scales, the Full-Body AI Agent framework represents a paradigm shift in how human biology is modeled and interpreted. While recent advances demonstrate its promise, several key challenges remain. These challenges encompass existing data availability, technical limitations, and ethical considerations. Addressing these obstacles is crucial to realizing the full potential of this integrative, multi-scale approach to biomedical research and clinical application.

### 6.1. Existing data and technical support

Despite significant advancements in computational biology and AI, several technical bottlenecks must be addressed to fully realize the potential of the Full-Body AI-Agent framework. A major challenge is integrating heterogeneous data. Biological data comes from diverse sources, including genomic sequences, proteomic profiles, clinical records, and imaging data. These data types often differ in format, resolution, and scale, making it difficult to integrate them seamlessly into a unified model. The challenge of standardizing and harmonizing these datasets is a critical hurdle for the framework. Another challenge is model interpretability. Deep learning models, particularly those used in complex biological simulations, often function as "black boxes," meaning it is difficult to understand how they generate predictions or insights. In biological research, it is essential that the model outputs be interpretable so that they can be validated, trusted, and used effectively. This calls for the development of explainable AI (XAI) techniques that can shed light on how models arrive at their conclusions, especially when these insights will inform clinical decisions or influence therapeutic strategies. Scalability is also a significant issue. Modeling human biology across multiple levels requires immense computational resources. Simulating the dynamic interactions between these layers in real time involves handling massive datasets and complex algorithms. Current computational models may not be sufficient to scale with the increasing complexity of biological systems. Developing more efficient and scalable algorithms will be necessary to overcome this barrier and ensure that the Full-Body AI-Agent can handle the high computational demands of multi-level biological simulations. Furthermore, data quality and noise are ongoing challenges. Biological datasets are often incomplete or noisy, due to sampling biases, technical limitations, or variations between experiments. The Full-Body AI-Agent will need to incorporate robust methods for cleaning and standardizing data to minimize these issues. Missing values, inconsistencies, and noise must be addressed to ensure that the model is built on high-quality data, which is crucial for producing accurate and reliable predictions. Finally, the refinement of computational models and algorithms is an ongoing process. Biological systems are highly dynamic and complex, often exhibiting non-linear relationships and feedback loops. Current AI models may not be able to accurately capture the full complexity of these systems, so there is a need for continuous improvement in the algorithms used by the Full-Body AI-Agent to better simulate these intricate biological processes.

## 6.2. Technical bottleneck and extensions

Currently, there is a significant lack of effective tools capable of characterizing the human body across all biological levels, especially at the levels of organ interactions and whole-body systems. To address these gaps, we have developed and proposed specialized methods and tools aiming at enhancing the capabilities of the Full-body AI Agent in future applications. One such innovation is Spatial-GWAS, a method we defined to enable a more comprehensive investigation of human systemic biology integrated into the Full-Body AI-Agent framework, which simultaneously considers brain spatial interactions and genetic association studies, improving the accuracy of identifying important genetic variations and elucidating their interactions.

## 6.3. Ethics and application limitations

The Full-Body AI-Agent framework raises several ethical concerns, primarily around data privacy and security. Protecting sensitive patient data, including genomic and clinical information, requires robust encryption, anonymization, and secure access protocols to ensure confidentiality and maintain patient privacy. Bias and fairness are also significant issues, as unrepresentative training data could lead to biased predictions and health disparities. Additionally, the framework must undergo clinical validation to ensure accuracy and reliability before being used in healthcare settings. Transparency and accountability are key, as AI-generated insights must complement human expertise with clear guidelines on decision-making responsibilities. Finally, effective regulatory oversight will be crucial to ensuring compliance with healthcare standards and ethical practices in real-world applications.

## 7. Discussion

The Full-Body AI-Agent framework, as a forward-thinking proposal, presents an innovative approach to modeling human biology through a network of specialized AI Agents. Designed to operate at multiple biological levels, this framework offers a novel lens through which to simulate and predict complex biological processes. This framework envisions an unprecedented integration of diverse biological data sources, including genetic data, chemical and physicochemical properties, cell biology data, physiological data, anatomical structures, and multi-scale imaging. By unifying these data resources, the Full-Body AI Agent has the potential to significantly expand the scope and resolution of biomedical research. Unlike domain-bounded systems that optimize for isolated workflow segments, the Full-Body AI-Agent operationalizes reasoning as an intrinsically dynamic, cross-scale process. This shift has implications beyond mere technical integration: it reframes how computational systems can emulate biological causality, not by aggregating static outputs, but by allowing hypotheses to evolve through iterative, bidirectional exchanges between mechanistic and systemic levels.

The Full-Body AI-Agent framework represents a paradigm shift in biomedical modeling, advancing from domain-specific optimization toward intrinsically dynamic, cross-scale reasoning. By embedding iterative, bidirectional exchanges between mechanistic and systemic levels, it reframes how computational systems emulate biological causality. This architectural shift has profound implications for how molecular discoveries, physiological simulations, and translational decisions can be aligned within a single inferential process.

Methodologically, the framework addresses two enduring weaknesses in biomedical AI. First, it mitigates the fidelity loss that occurs when molecular- or organ-level findings are extrapolated without systemic context, ensuring that explanatory models remain both mechanistically sound and physiologically plausible. Second, it expands inference from targeted problem solving to whole system hypothesis generation, enabling the capture of emergent properties, inter organ interactions, and polypharmacological effects, phenomena that are often inaccessible to reductionist approaches.

A key advantage lies in its ability to situate experimental platforms such as organoids, organ-on-chip systems, and in vivo studies within a coherent computational reasoning loop. Rather than displacing these methods, the Full-Body AI-Agent can strategically prioritize data acquisition, identify high-value experimental validations, and reduce redundant testing. This coupling of in silico and wet-lab pipelines could help alleviate the persistent bottlenecks and high attrition rates in drug development, a challenge well-documented in both academic and industrial settings.

Realizing this potential, however, requires overcoming substantial technical barriers. Data fragmentation remains a critical constraint: biological datasets vary widely in resolution, scale, and modality, impeding seamless integration. Standardization and harmonization pipelines, for both experimental outputs and public databases, will be essential for enabling multi-agent collaboration. Furthermore, the temporal alignment of heterogeneous datasets, particularly for dynamic physiological processes, is still a largely unsolved problem. On the computational side, organism-scale simulations with sufficient fidelity remain resource-intensive, and interpretability of multi-scale causal chains remains a major barrier to clinical adoption.

Advances in experimental biology will play a decisive role. High-resolution, multi-dimensional modalities such as single-cell and spatial omics, multi-scale imaging, and continuous physiological monitoring are essential to populate the framework's reasoning layers with relevant, time-resolved data. Parallel developments in AI, including deep learning, reinforcement learning, and LLMs, will further enhance the system's capacity to process diverse data types and integrate structured and unstructured biomedical knowledge. Yet, the "black box" nature of many deep models poses challenges for explainability. Building trust, particularly in clinical settings, will require interpretability mechanisms that transparently map AI-generated inferences to established biomedical principles.

Ethical and governance considerations are equally critical. The integration of patient-specific, multi-modal data amplifies privacy, security, and bias concerns. Compliance with frameworks such as GDPR and HIPAA must be embedded by design, alongside robust encryption, anonymization, and bias-mitigation protocols. Without careful attention to data diversity, AI-driven recommendations risk perpetuating or exacerbating health inequities. Moreover, the framework should be positioned as an augmentative tool, with final decision-making authority remaining firmly in the hands of medical professionals.

Looking forward, clinical translation represents the most promising but also most demanding frontier. In personalized medicine, the framework could model individual biology at multiple scales to optimize treatment selection, dosing, and timing. In drug discovery, it could unify molecular docking, organoid simulation, and whole-body pharmacodynamics into a single reasoning cycle, reducing both cost and time-to-clinic. However, these applications will require rigorous, prospective validation to ensure that computational predictions reliably translate into therapeutic benefit.

Ultimately, the Full-Body AI-Agent should be seen less as a fixed technological artifact and more as an evolving paradigm, one capable of integrating disparate data streams across biological levels, harmonizing experimental and computational biology through task orchestration, and continuously refining models via iterative evidence-based loops. If its technical, methodological, and ethical challenges can be addressed, it holds the potential to deliver unprecedented insights into human biology, enabling more precise interventions, accelerating therapeutic innovation, and deepening our understanding of the systemic nature of health and disease.

**Funding**

Wang, Luo, Fan, Hu, and Yu were funded by National Institutes of Health (NIH R01GM123037, U01AR069395, R01CA241930), and National Science Foundation (NSF 2217515, 2326879); Liu, Wen, Zhao, Li, Zhou were partially supported by National Institutes of Health (R01LM014156, R01CA241930, R01GM153822), CPRIT (RP250043), National Science Foundation (NSF2217515 and NSF2326879), and Dr. & Mrs. Carl V. Vartian Professorship.

**Figure legends**

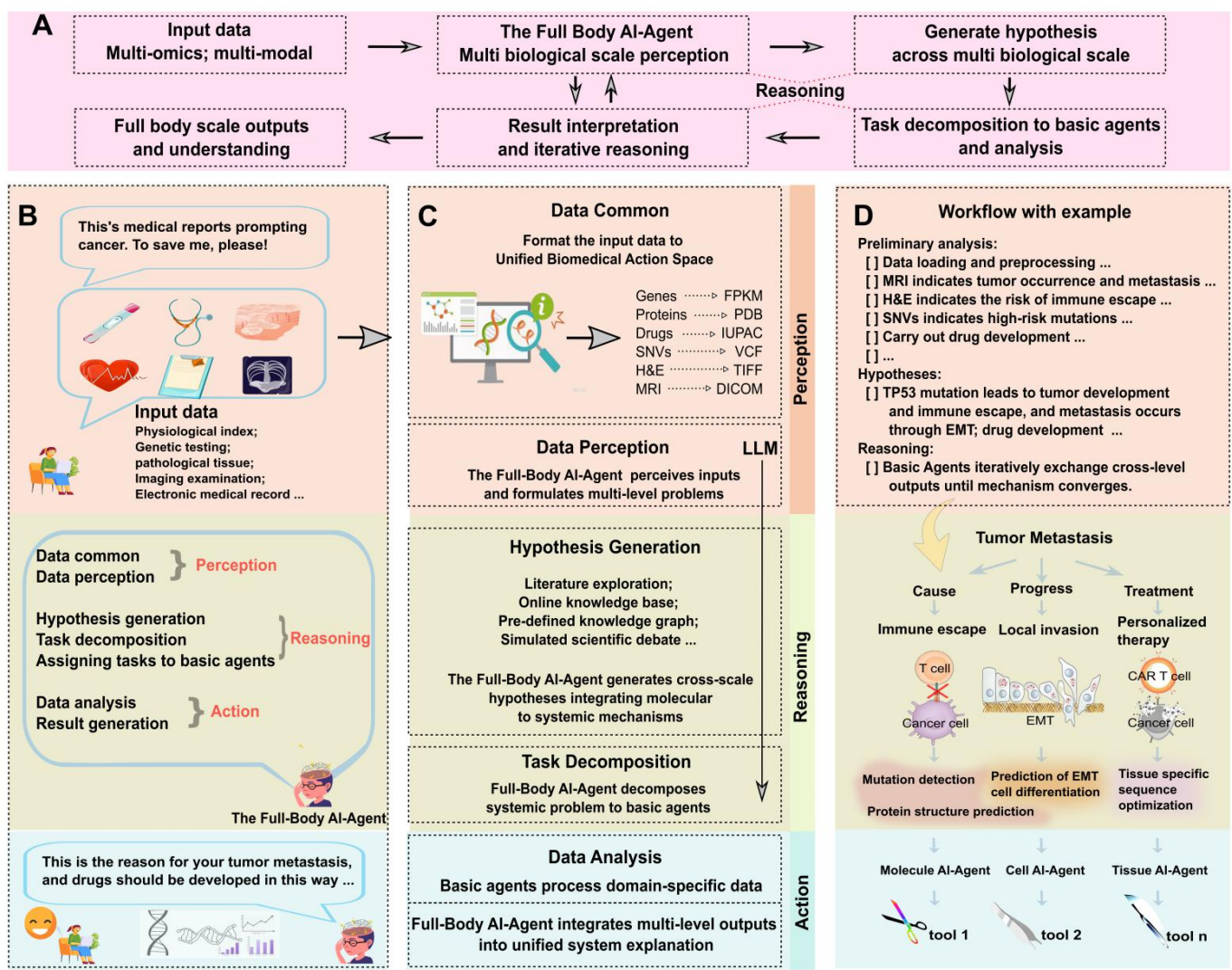

Figure 1. **Schematic of the Full-Body AI Agent framework for systemic human biology analysis and therapeutic Insight generation.** (A). presents the core cycle, where multi-modal and multi-omics data are input, the Full-Body AI Agent as a supervisor perceives and preprocesses data to generate cross-scale hypotheses, decomposes tasks for basic agents, which execute level - specific analyses, and the supervisor integrates results for systemic output; (B). shows patient interaction, with diverse clinical inputs like physiological data and EHRs entering the workflow; (C). highlights LLM - driven reasoning, running through data perception, hypothesis generation and task decomposition via literature and knowledge bases; (D). uses a TP53 mutation - related oncology example, where the Full-Body AI Agent perceives data, decomposes tasks for basic agents to do mutation detection, protein prediction etc., and guides drug design, enabling dynamic human biology modeling for precision medicine.

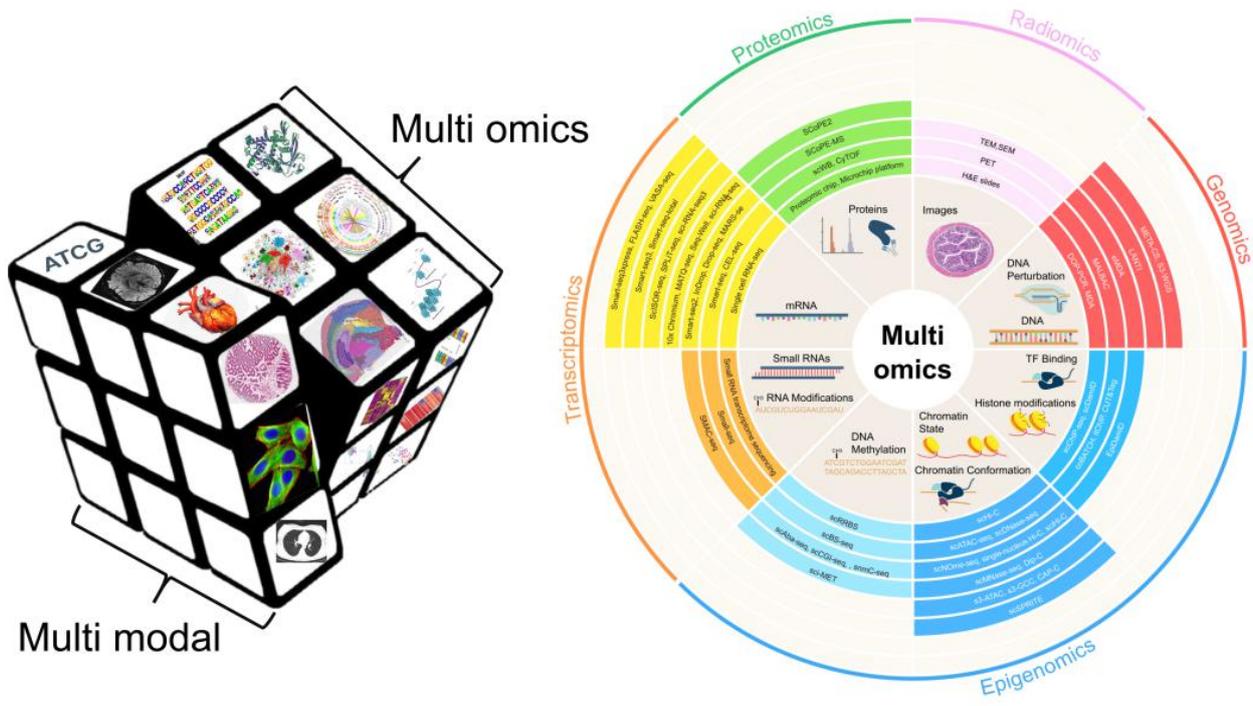

Figure 2. **Overview of multi-modal and multi-omics of biomedical data types.** Complex biological data involve genomics, epigenomics, transcriptomics, proteomics and imaging, and are stored in the form of numerical values, text, images, etc.

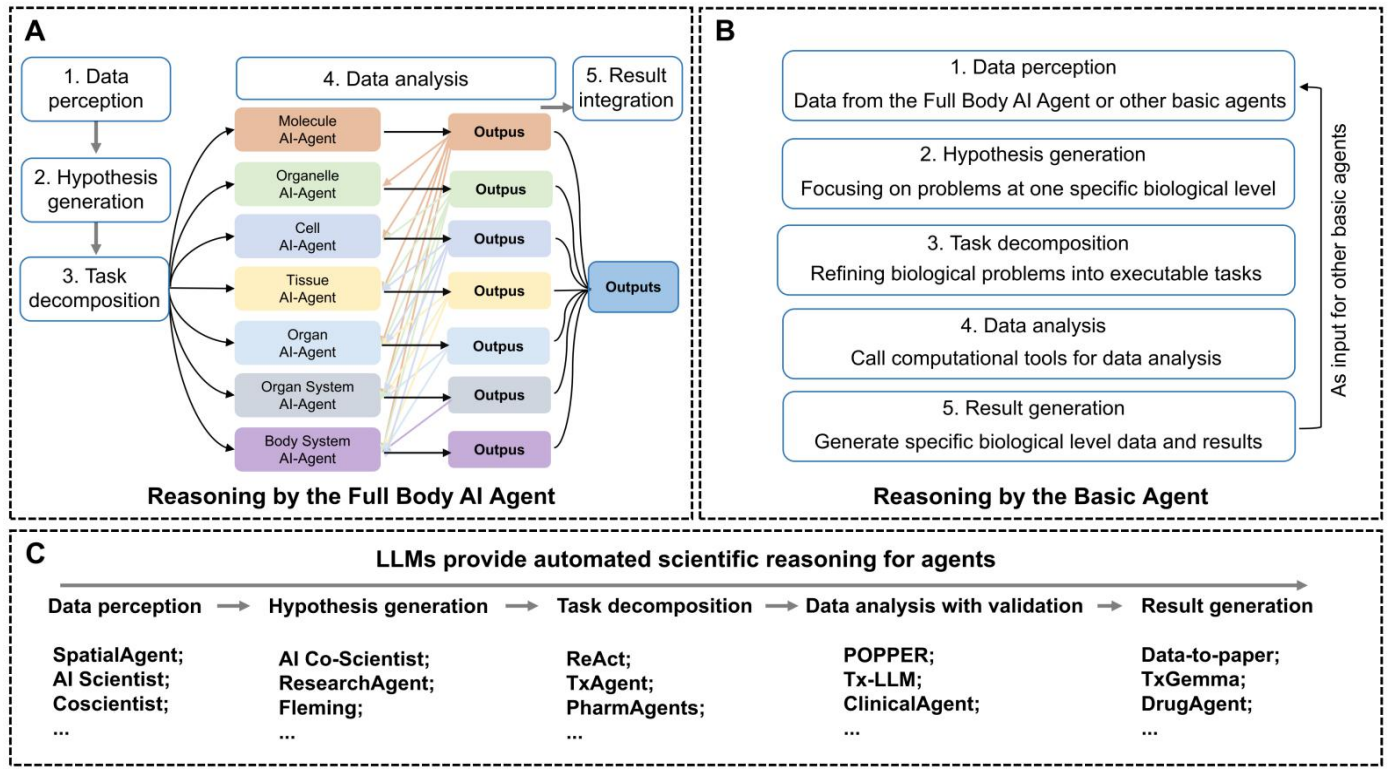

**Figure 3. Hierarchical reasoning framework of Full-Body and basic AI Agents for multi-scale biological discovery.** (A). The Full Body AI Agent executes a complete scientific reasoning loop, starting from data perception and gradually decomposing hypothesis generation tasks into data analysis, until the

results are integrated and coordinated with multi-level agents such as molecular and cell to output a unified conclusion. (B). After receiving data, the Basic agent performs hypothesis generation, task decomposition, and tool analysis within a single biological level, and outputs specific hierarchical results. (C) The LLM toolchain supports the complete inference process, among which TxAgent, AI Scientist, and PharmAgents support the entire process from data perception, hypothesis generation, task decomposition, and data analysis to generate results of systems biology.

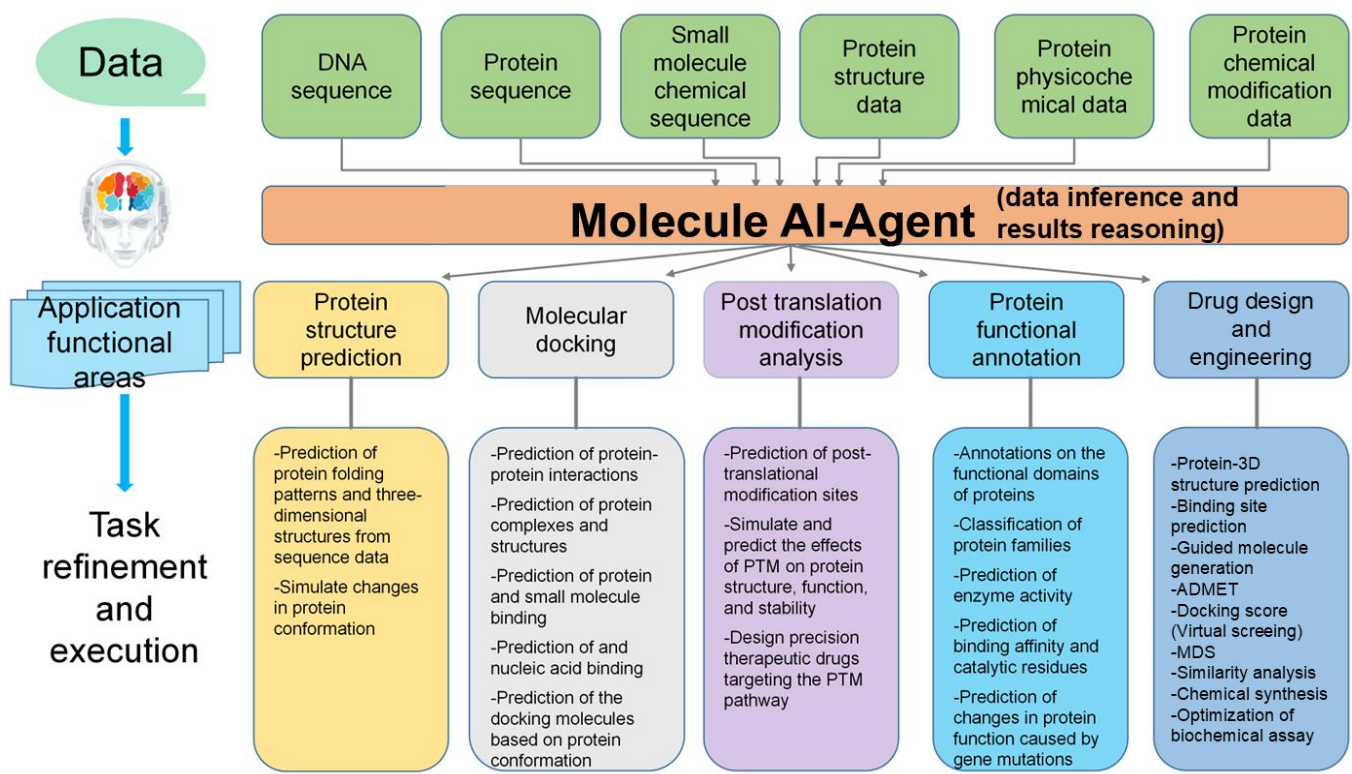

Figure 4. **Predefined detachable biological tasks of the Molecule AI Agent.** It presents the data types perceived by the Molecule AI Agent, the application functional areas of biological hypotheses, as well as the detailed decomposition and execution of tasks.

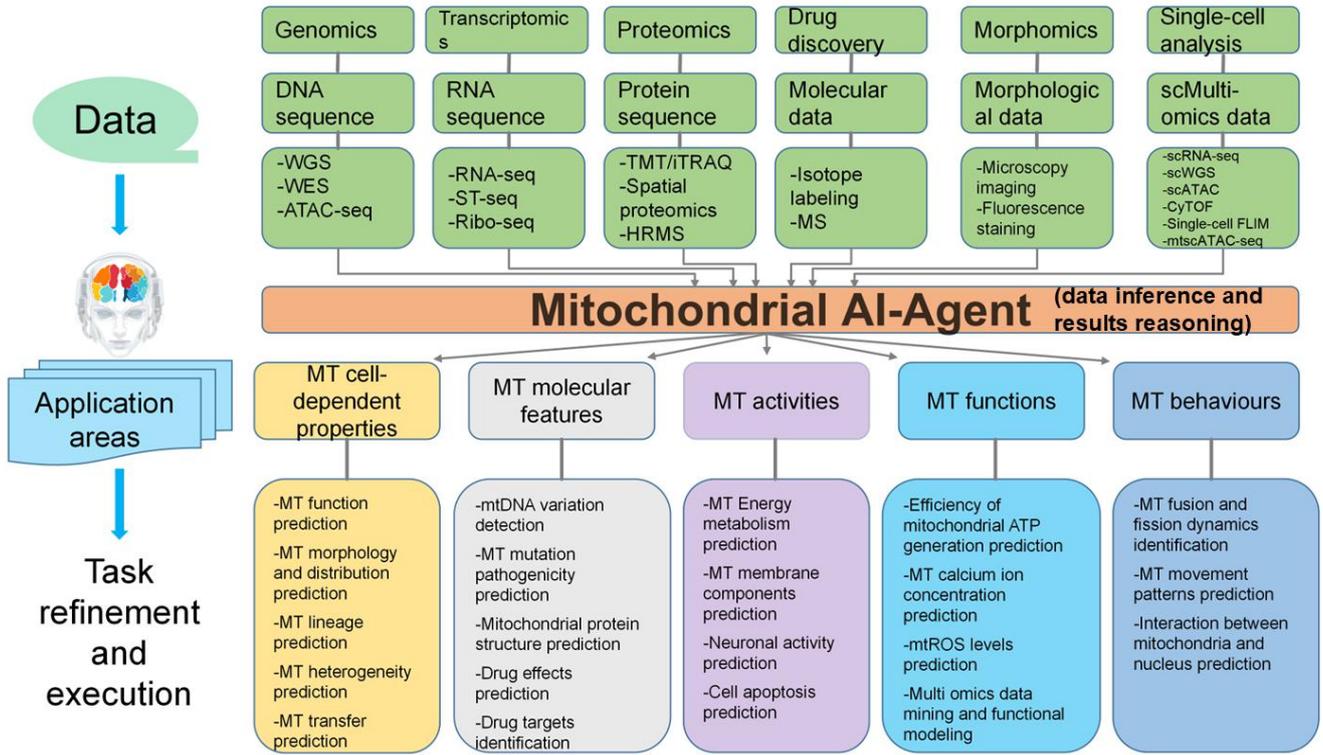

Figure 5. **Predefined detachable biological tasks of the Mitochondrial AI-Agent.** It presents the data types perceived by the Mitochondrial AI Agent, the application functional areas of biological hypotheses, as well as the detailed decomposition and execution of tasks.

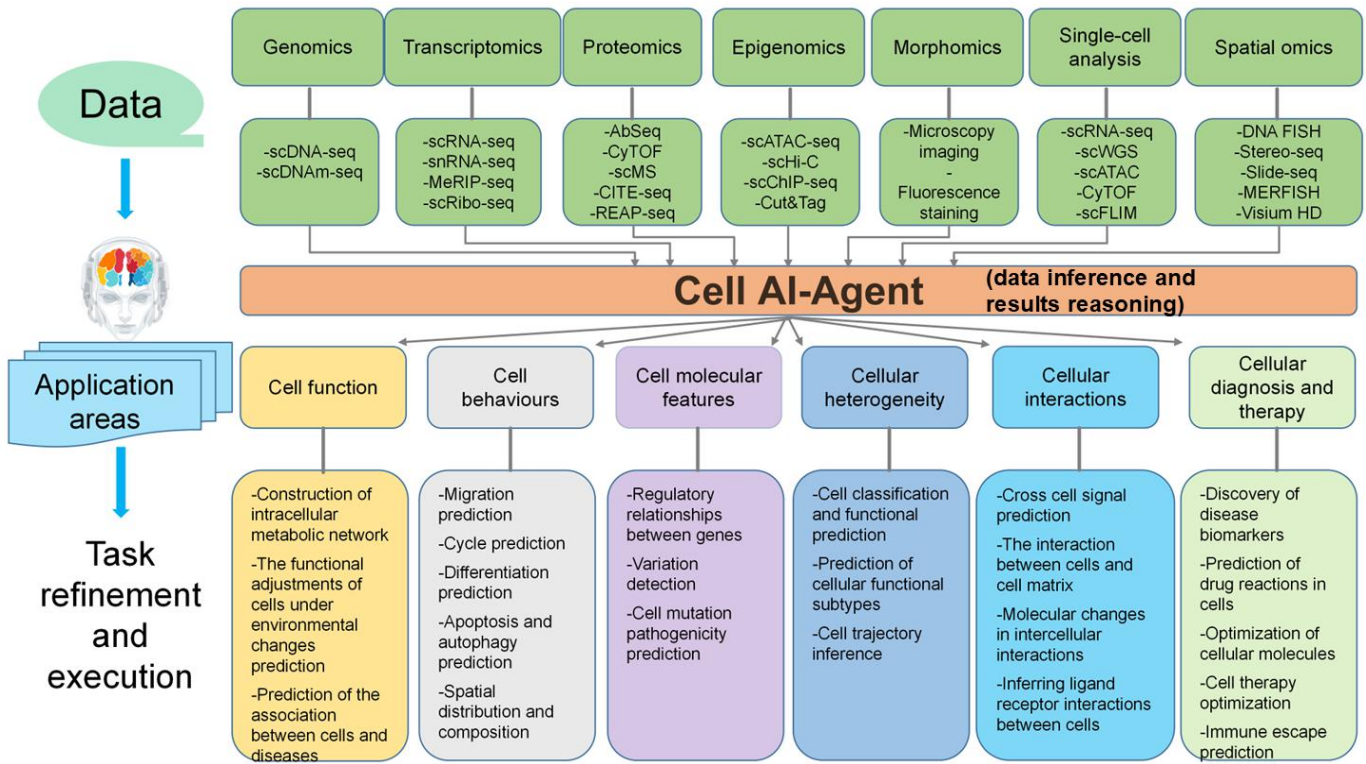

Figure 6. **The predefined detachable biological tasks assigned to the Cell AI Agent.** It presents the data types perceived by Cell AI Agent, the application functional areas of biological hypotheses, as well as the detailed decomposition and execution of tasks.

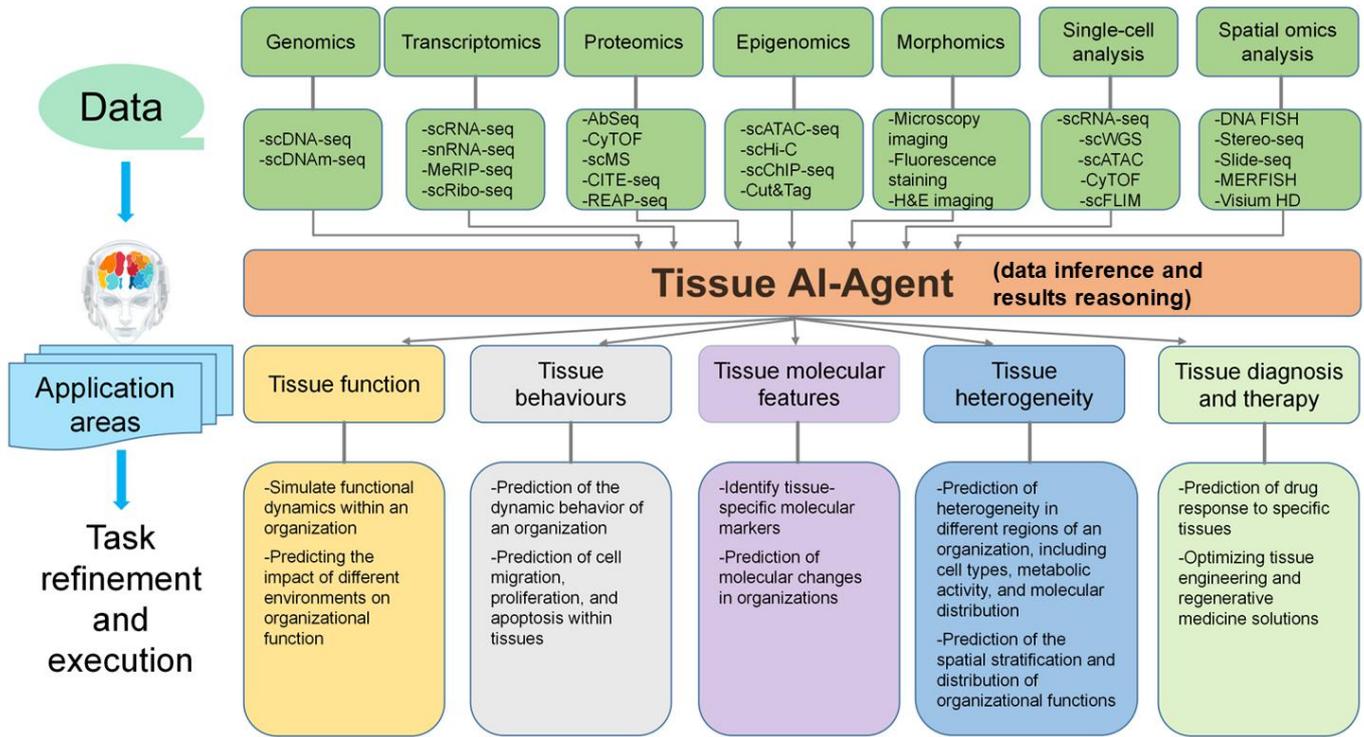

Figure 7. **The predefined detachable biological tasks for the Tissue AI Agent.** It presents the data types perceived by Tissue AI Agent, the application functional areas of biological hypotheses, as well as the detailed decomposition and execution of tasks.

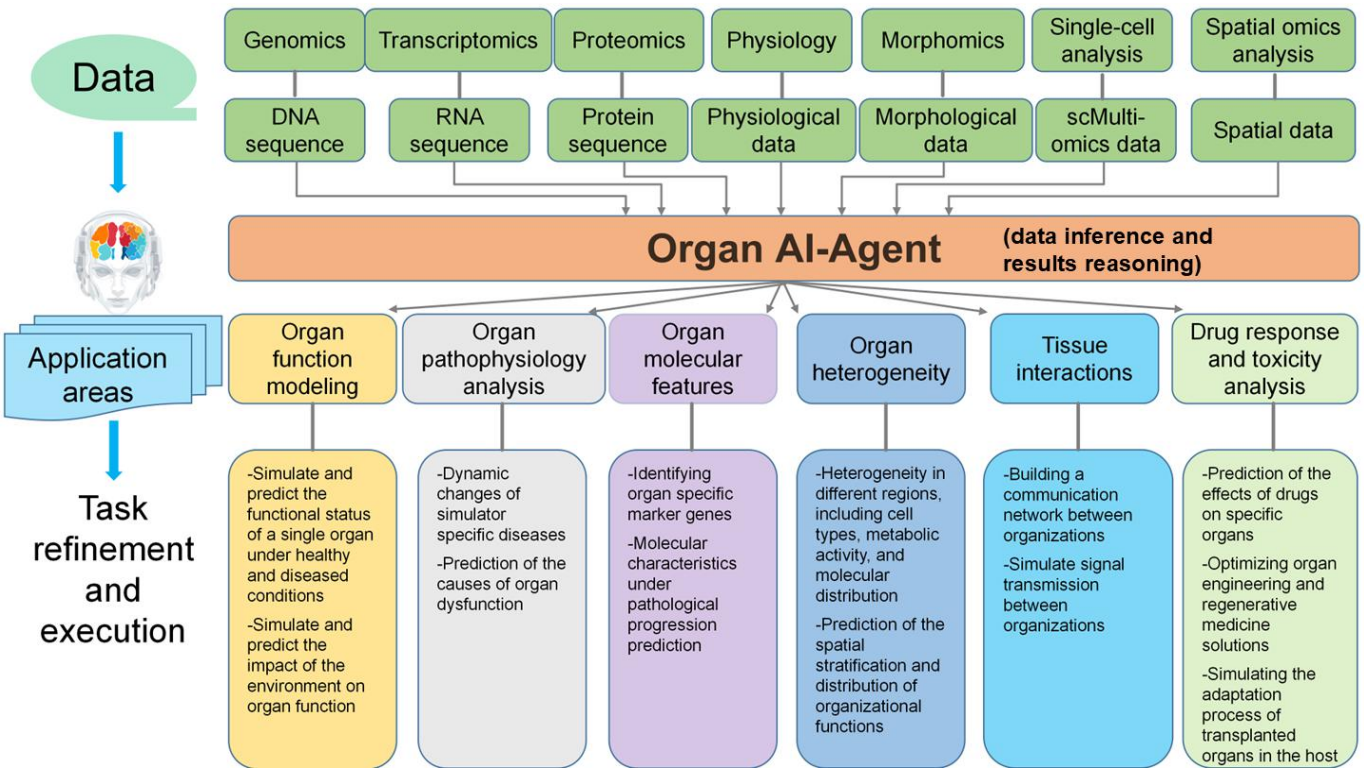

Figure 8. **The predefined detachable biological tasks performed by the Organ AI Agent.** It presents the data types perceived by Organ AI Agent, the application functional areas of biological hypotheses, as well as the detailed decomposition and execution of tasks.

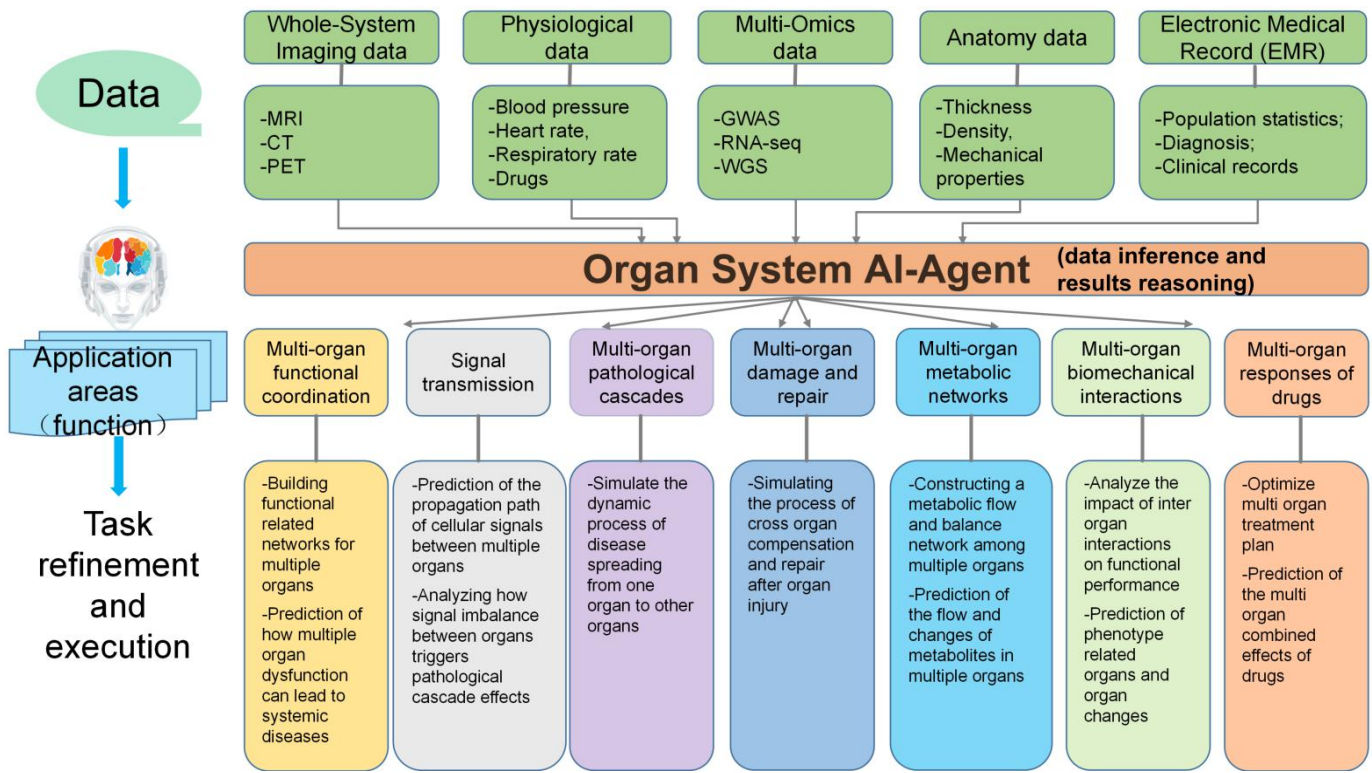

Figure 9. **Predefined detachable biological tasks for the Organ System AI Agent.** It presents the data types perceived by Organ System AI Agent, the application functional areas of biological hypotheses, as well as the detailed decomposition and execution of tasks.

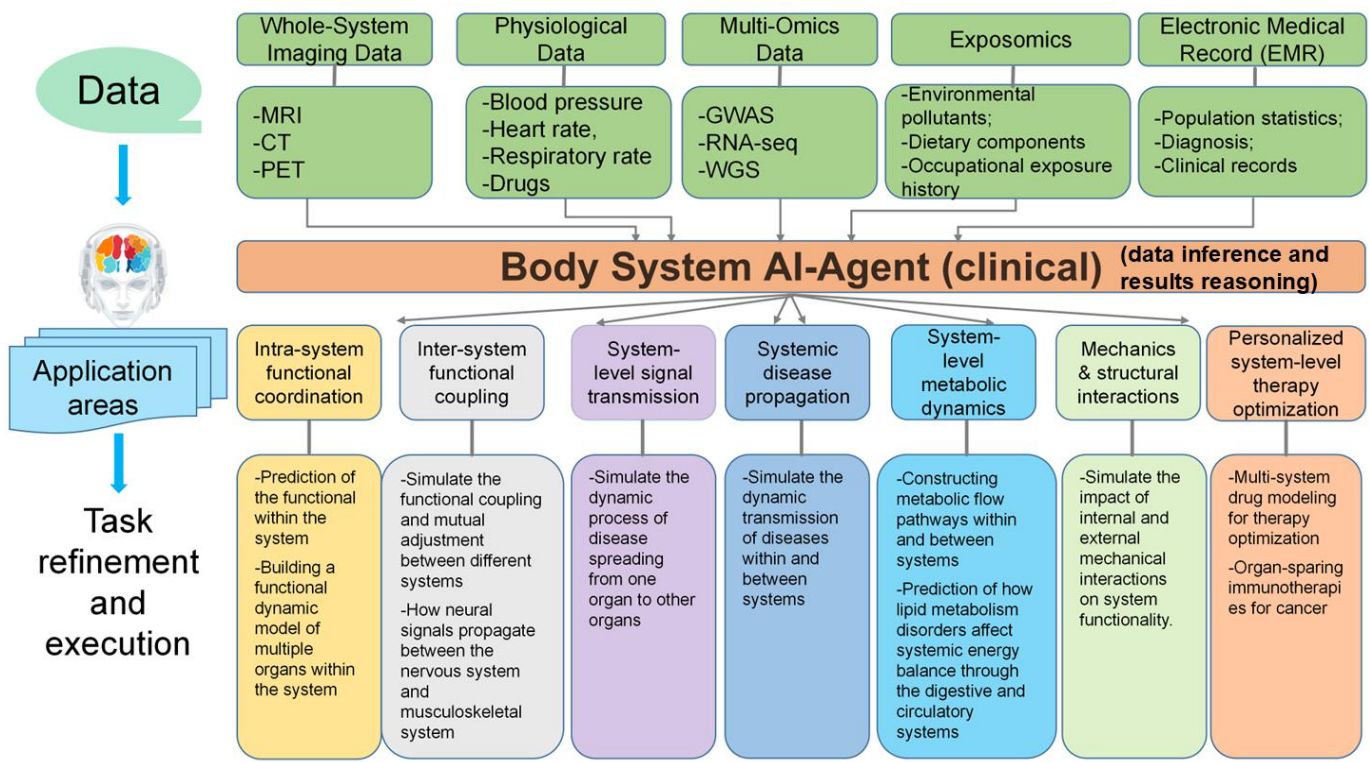

Figure 10. **Predefined detachable biological tasks for the Organ System AI Agent.** It presents the data types perceived by Body System AI Agent, the application functional areas of biological hypotheses, as well as the detailed decomposition and execution of tasks.

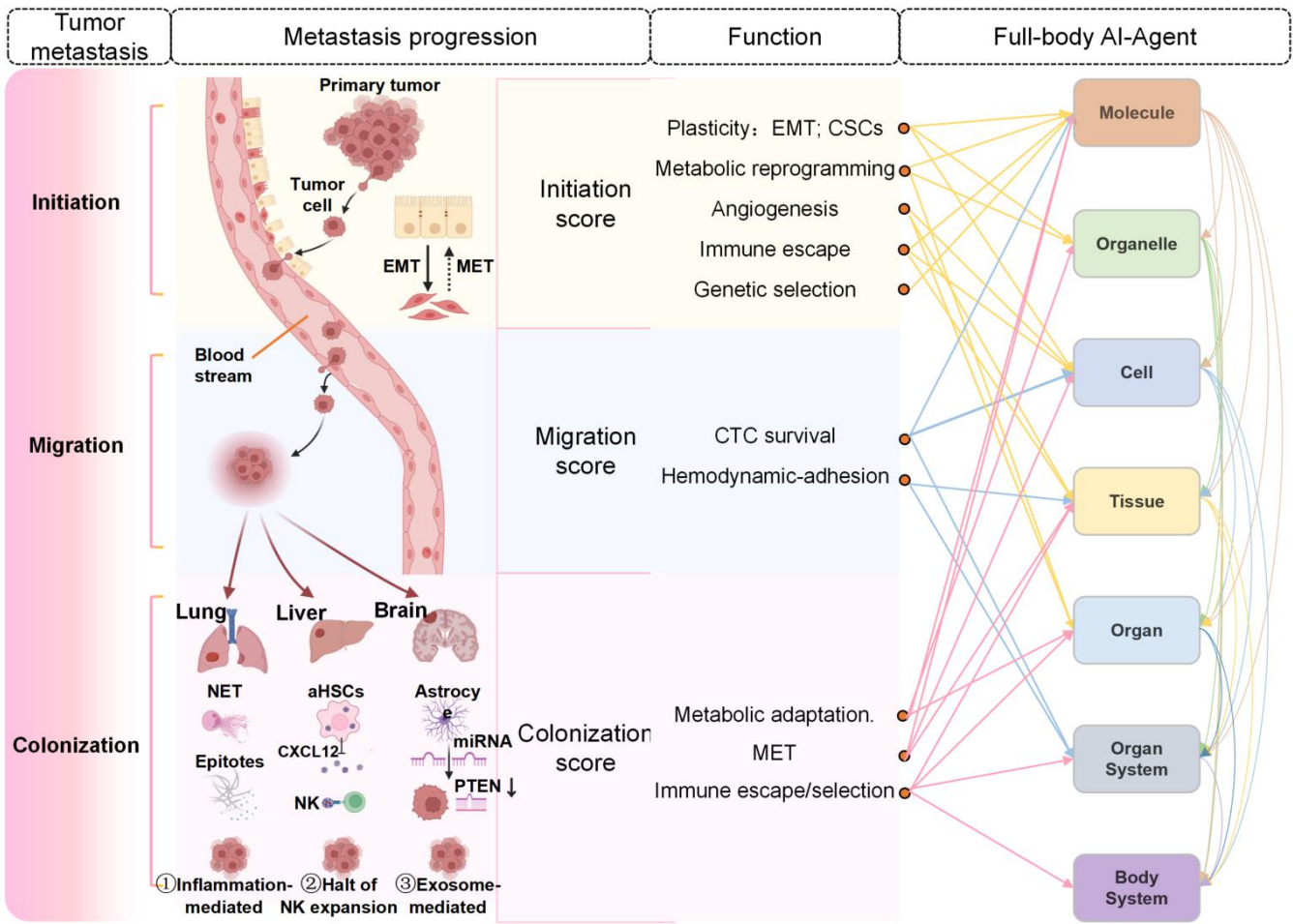

Figure 11. **Metastasis AI Agent, a three-phase metastasis scoring framework leveraging Full-Body AI-Agent system**. (A). The metastatic cascade unfolds in three phases: primary tumor invasion, vascular dissemination, and distant colonization. (B). Organ-specific colonization mechanisms: lung trapping by NETs/CXCL12, liver dormancy enforced by stellate cell-mediated NK suppression, and brain outgrowth primed by astrocyte exosomal PTEN loss. (C). Quantitative scoring of each phase (Initiation, Dissemination, and Colonization) captures the key biological hurdles. (D). Integration of multi-scale, multi-modal features, from molecular to systemic, enables comprehensive evaluation of metastatic potential.

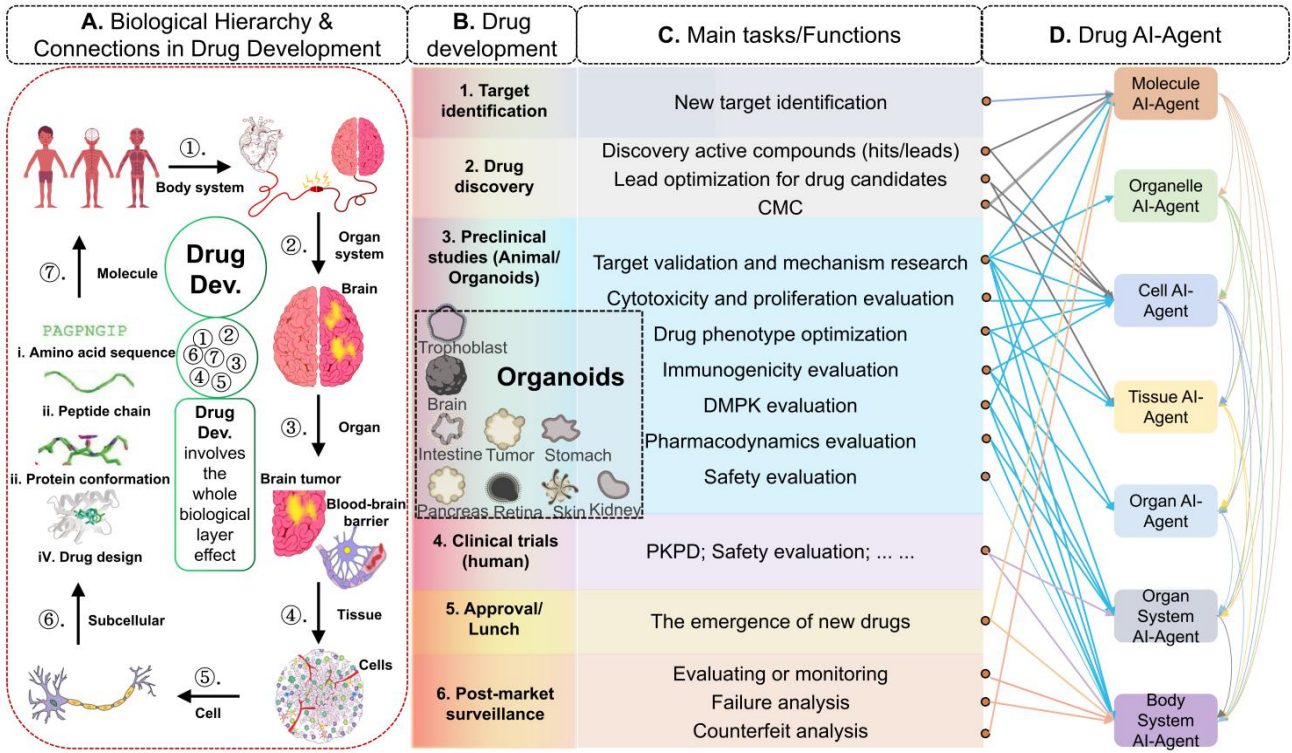

Figure 12. **Integration of biological hierarchy, drug development workflow, and drug AI-Agent architecture.** This figure illustrates the multi-scale integration of biological organization (A) into the canonical drug development pipeline (B), with organoids highlighted as key tools in preclinical research. Panel (C) outlines the major tasks and biological functions required across development stages, while Panel (D) shows how each foundational AI-Agent within the Full-Body AI-Agent framework contributes to these tasks. The Drug AI-Agent functions as an integrative layer that synthesizes outputs across all levels to support system-wide prediction, optimization, and feedback in drug development.

# Table Notes

| Multi-Agents systems | OriGene | Robin | Biomni | MAC-doctor | PharmaSwarm | AI Co-Scientist | Full-Body AI-Agent |
|---|---|---|---|---|---|---|---|
| General Purpose | Discover novel therapeutic targets via self-evolving reasoning | Automate closed-loop discovery with multi-agent lab simulation | Biomedical agent for 25 subfields using code-native reasoning | AI-assisted diagnostics with patient-specific data | Integrate omics, literature and market for drug discovery | Multi-agent workflow from problem to results | Reasoning across molecular to physiological levels to address systemic biological questions |
| Components | Coordinator, Planning, Reasoning, Critic, Reporting Agent | Crow, Falcon, Finch Agent | Retrieving, Reasoning, CodeAct Agent | Supervisor Agent, Doctor Agents | Terrain2Drug, Paper2Drug, Market2Drug Agent | Generation, Reflection, Ranking, Evolution, Meta-review Agent | Basic Agents and Full-Body AI-Agent |
| Input Perception | Coordinator parses user query | User provides disease; triggers Crow/Falcon/Finch | Retrieving fetches multi-source data | Supervisor parses problem; assigns to Doctors | Detects disease/target; triggers domain agents | Generation parses goals; calls tools | Full-Body AI-Agent perceives inputs and formulates multi-level problems |
| Hypothesis Generation | Coordinator decomposes; Reasoning proposes targets | Crow mines literature for candidates | Retrieving gathers datasets | Supervisor formulates hypotheses | Domain agents generate targets from omics, papers and market | Generation proposes from multi-source evidence | Full-Body AI-Agent generates cross-scale hypotheses integrating molecular to systemic mechanisms |
| Task Decomposition | Coordinator assigns subtasks | Falcon plans workflow, assigns agents | Reasoning embeds planning | Supervisor delegates tasks | System sequences agents | Supervisor coordinates generation, evaluation and reflection Agents | Full-Body AI-Agent decomposes systemic problem to basic agents |
| Data Analysis | Planning selects tools, processes data | Finch synthesizes causal chains | Reasoning processes data via 150+ tools | Doctors analyze images, labs and history clinical information | Domain agents perform enrichment, literature and market analytics | Gemini 2.0 and Generation analyze multimodal data | Basic agents process domain-specific data |
| Reasoning | Reasoning integrates multimodal outputs | Finch outputs refine reasoning | Reasoning evaluates mechanisms | Doctors integrate analyses | Cross-agent scoring and prioritization | Reflection evaluates and refines | Basic Agents iteratively exchange cross-level outputs until mechanism converges |
| Result Interpretation | Critic and Reporting validate and format | Integrates findings; ranks hypotheses | CodeAct converts results to workflows | Supervisor finalizes report | Scoring ranks outputs | Ranking and Evolution rank; diversify | Full-Body AI-Agent integrates multi-level outputs into unified system explanation |
| Output | Reporting produces final report | Designs, ranked drugs and visualizations | CodeAct outputs actions | Supervisor compiles summary | Report from ranked results | Meta-review integrates literature and evidence | Full-Body AI-Agent generates full-body maps, simulations and ranked strategies |
| Data & Tool Access | Accesses 500+ biomedical databases, pathway tools | Accesses literature via LLMs with PaperQA2 and local file repositories | Accesses 150 specialized tools, 105 software packages, and 59 biomedical databases | Accesses embedded medical knowledge base via Doctor and Supervisor Agents | Accesses omics data, literature, drug databases, and knowledge graphs via GET module | Accesses online literature databases, scientific web content, and local documents | Accesses to multi-scale biomedical data and computational tools |
| Biological Cases | Breast cancer: novel targets validated in organoids | dAMD: automated review, hypothesis, validation within anti-FGF2 candidate | Colorectal cancer: automated single-cell immune profiling | Sepsis: early shock detection; tailored antibiotics | Pancreatic cancer: repurposed glycolysis inhibitor validated preclinically | Neurodegeneration: iterative tau pathway modeling; intervention model | Systemic disease modeling (lung cancer metastasis, whole-body drug development) |

Table 1. **Comparative of multi-AI Agent systems in biomedical research.** The table compares the differentiation strategies of OriGene, Robin, Biomeni, MAC-doctor, PharmaSwarm, and AI Co Scientist with the Full Body AI Agent system, including multi-agent systems composed of different agents with different general purposes. And their reasoning process and data input and output.

**Supplementary Data**

Supplementary Table 1. **Multi-omics and Multi-modal Data Mapping to Biological Hierarchies.** Detailed list of standardized methods, data types, and common data formats for omics or modal biological data, spanning 7 levels and mapped to basic agents. Can be downloaded from

https://figshare.com/articles/dataset/Supplement_table1-Multi-omics_and_Multi-modal_Data_Mapping_to_Biological_Hierarchies/29974639?file=57381001